\documentclass[conference]{IEEEtran}

\usepackage{cite}
\usepackage{url}
\usepackage{graphicx}
\usepackage{subfigure}
\usepackage[table,xcdraw]{xcolor}
\usepackage{adjustbox}
\usepackage{array}
\usepackage{multirow}
\usepackage{makecell}
\usepackage{balance}
\usepackage{amsmath}
\newcolumntype{L}[1]{>{\raggedright\let\newline\\\arraybackslash\hspace{0pt}}m{#1}}
\newcolumntype{C}[1]{>{\centering\let\newline\\\arraybackslash\hspace{0pt}}m{#1}}
\newcolumntype{R}[1]{>{\raggedleft\let\newline\\\arraybackslash\hspace{0pt}}m{#1}}

\newcommand{\smallsim}{\smallsym{\mathrel}{\sim}}

\makeatletter
\newcommand{\smallsym}[2]{#1{\mathpalette\make@small@sym{#2}}}
\newcommand{\make@small@sym}[2]{%
	\vcenter{\hbox{$\m@th\downgrade@style#1#2$}}%
}
\newcommand{\downgrade@style}[1]{%
	\ifx#1\displaystyle\scriptstyle\else
	\ifx#1\textstyle\scriptstyle\else
	\scriptscriptstyle
	\fi\fi
}
\makeatother

\begin{document}
\title{Estimating Residential Broadband Capacity \\using Big Data from M-Lab}

\author{\IEEEauthorblockN{Xiaohong Deng, Yun Feng, Hassan Habibi Gharakheili and Vijay Sivaraman}
			\IEEEauthorblockA{School of Electrical Engineering and Telecommunications, University of New South Wales, Sydney, Australia\\
			Emails:\{xiaohong.deng, yun.feng, h.habibi, vijay\}@unsw.edu.au}
		}

\maketitle


\begin{abstract}
Knowing residential broadband capacity profiles across a population is of interest to both consumers and regulators who want to compare or audit performance of various broadband service offerings. Unfortunately, extracting broadband capacity from speed tests in public datasets like M-Lab is challenging because tests are indexed by client IP address which can be dynamic and/or obfuscated by NAT, and variable network conditions can affect measurements. This paper presents the first systematic effort to isolate households and extract their broadband capacity using 63 million speed test measurements recorded over a 12 month period in the M-Lab dataset. We first identify a key parameter, the correlation between measured speed and congestion count for a specific client IP address, as an indicator of whether the IP address represents a single house, or a plurality of houses that may be dynamically sharing addresses or be aggregated behind a NAT. We then validate our approach by comparing to ground truth taken from a few known houses, and at larger scale by checking internal consistency across ISPs and across months. Lastly, we present results that isolate households and estimate their broadband capacity based on measured data, and additionally reveal insights into the prevalence of NAT and variations in service capacity tiers across ISPs.
\end{abstract}
\IEEEpeerreviewmaketitle

\section{Introduction}
Broadband performance continues to be of great interest to consumers and regulators: Netflix publishes a monthly ISP speed index \cite{Netflix} that ranks ISPs based on their measured prime time Netflix performance, and Youtube graphs for each ISP the break-down of video streaming quality (low vs standard vs high definition) by time-of-day averaged over a 30-day period \cite{Youtube}; the FCC in the US \cite{FCC} directs consumers to various speed test tools to make their own assessment, and the ACCC in Australia \cite{ACCC} is running a pilot program to instrument volunteers' homes with hardware probes to measure their connection speeds.

While the aggregated broadband service performance measures provided by large content providers and regulatory bodies are undoubtedly useful to consumers, the raw data underpinning these results are not openly available to the public. This seriously limits the ability of researchers to evaluate the importance of various factors that can underly differences in broadband performance amongst service providers, such as composition of metropolitan versus rural customers, proximity of content caches, network oversubscription factor used by the service provider, and even client OS and TCP versions. Some of these factors (like client OS) are outside the control of service providers, and indeed at least one large ISP claims that it ranks poorly on the Netflix index because it serves more rural customers who have lower connection capacities as well as older operating systems in their host clients.

In response to the need for open data on broadband performance, the research community initiated the M-Lab measurement platform \cite{M-Lab} that contains a wealth of tools to measure and record Internet performance. Specifically, it contains the Network Diagnostic Test (NDT) that records a rich set of attributes (pertaining to server, client, and network) for each broadband speed-test it performs, made available as open data. Tens of millions of speed-test measurements are collected world-wide every month, providing a rich data set for researchers to explore the various aspects influencing broadband performance.

There are however some challenges in using M-Lab data for studying broadband performance by household. The speed-test results are recorded by client (public) IP address, which may not map to a house. One IP address could represent many houses if the service provider uses NAT, or each house may have obtained varying IP addresses over time due to dynamic lease. The absence of any ground truth with regard to the mapping of IP addresses to houses makes it exceedingly difficult to draw meaningful inferences with regard to broadband capacity on a household basis.

The aim of this paper is to develop a method to make M-Lab data more useful by mapping measurements with households, so that meaningful performance comparisons across households can be conducted. Our specific contributions are as follows:
\begin{itemize}
	\item \textbf{First}, we identify a key attribute, namely the correlation between measured speed and congestion count, as an indicator of  whether an IP address corresponds to a single house, or a plurality of houses sharing dynamic address lease or aggregated behind a NAT.
	\item \textbf{Second}, we validate our method for detecting if an IP address belongs to a single household, by comparing to ground truth at small scale for specific houses, and by evaluating consistency at multiple time scales using large-scale M-Lab data.
	\item \textbf{Third}, we apply filtering to eliminate data points with IP addresses corresponding to multiple homes, as well as measurement outliers, to estimate the broadband capacity of individual households, and present insights into the distribution of access capacities for ISPs of various sizes.  
\end{itemize}
Our work provides an essential foundation for other researchers to use and interpret M-Lab data at a household (rather than an individual IP) level, enabling new insights that were not feasible before. Our code will be made openly available to the research community upon acceptance of this paper.  

The rest of this paper is organized as follows: \S\ref{sec:prior} recaps prior work in broadband performance, and gives relevant background on M-Lab dataset. In \S\ref{sec:isolation} we describe our method to isolate measurements of single households in M-Lab data. Validation of our method at small and large scales are discussed in \S\ref{sec:validation}, while in \S\ref{sec:insights} we apply our method to estimate broadband capacity and draw insights into variation of service capacity-tiers across ISPs. The paper is concluded in \S\ref{sec:concl}.

%\begin{itemize}
%	\item Why it is important to know access capacity of households (can compare broadband performance)?
%	\item Why it is challenging (confounders: NAT, IP address changes, outliers, etc.)? (one IP address could represent many houses due to NAT, or each house may have different IP addresses due to dynamic lease);
%	\item Absence of ground truth; 
%	\item Our approach and contributions. 
%	\item \textbf{First}, we argue that the correlation between measured speed and congestion count is a good indicator of  whether or not an IP address corresponds to a single household, and validate how the correlation distributions vary across ISPs of different size.
%	\item \textbf{Then}, we develop a method to  identify if an IP address corresponds to a single household, based on the consistency of the correlation at multiple time scales, allowing us to filter out data points with IP addresses that aggregate multiple houses behind NAT or get reassigned across households.
%	\item \textbf{Finally}, we apply outlier detection methods to estimate the broadband capacity of households, and show that our method corrects the distribution of access capacities, particularly for smaller ISPs that typically own smaller IP blocks.  
%	
%	Our method allows MLab data to be isolated and characterized by household, allowing the research community to make better inferences at the household level.  
%
%	
%\end{itemize}

\begin{figure*}[!t]
	\begin{center}
		\mbox{
			\subfigure[Negative correlation (Cox, 458 tests from 98.174.39.22) -- high speed during un-congested period, and low speed during fairly congested period.]{
				{\includegraphics[width=0.485\textwidth,height=0.27\textheight]{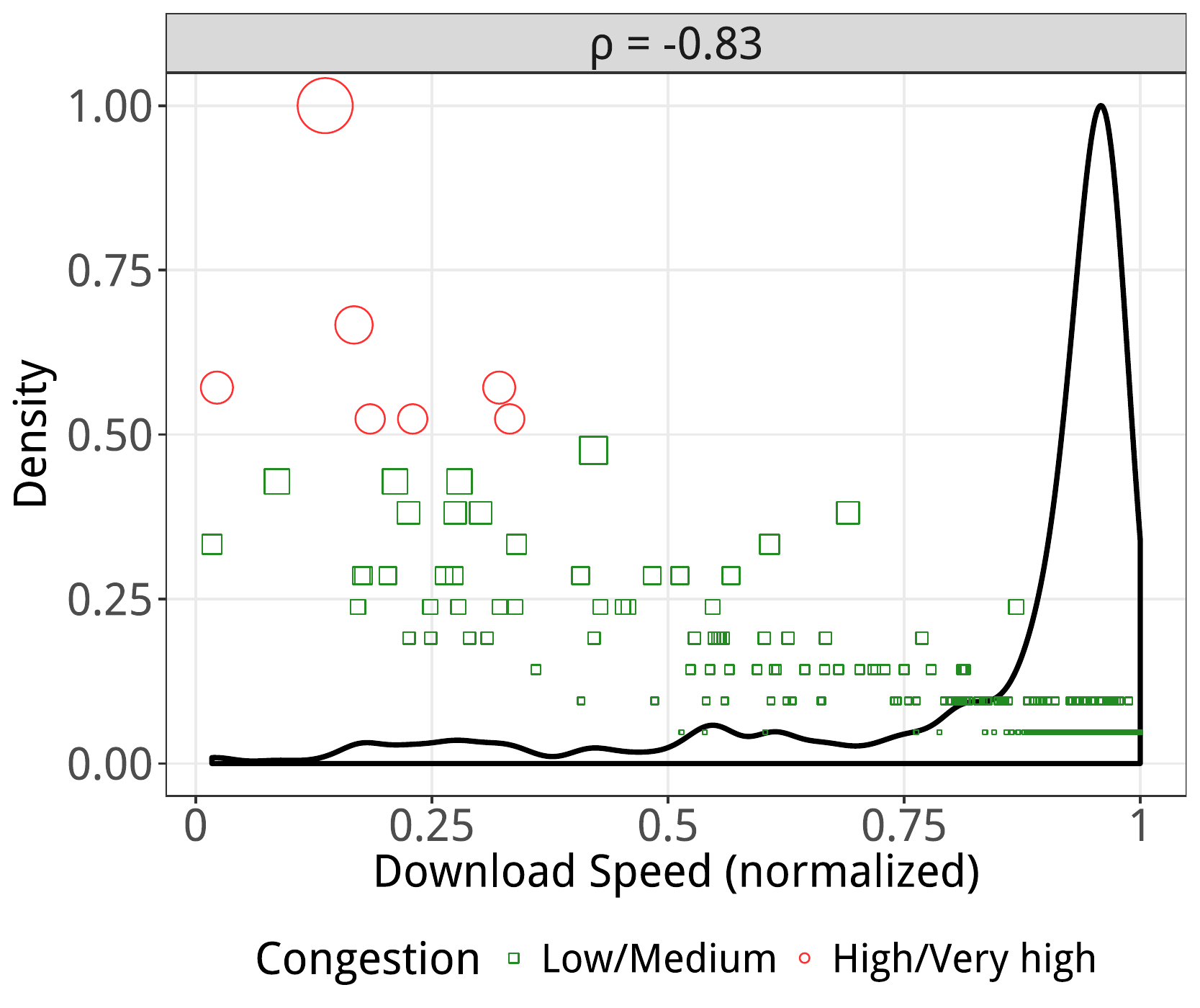}}\quad
				\label{fig:NegCorrAgg}
			}
			\hspace{0mm}
			\subfigure[Positive correlation (City of Thomasville Utilities), 896 tests from 64.39.155.194) -- high speed even during highly congested period, and low speed even during uncontested period.]{
				{\includegraphics[width=0.485\textwidth,height=0.27\textheight]{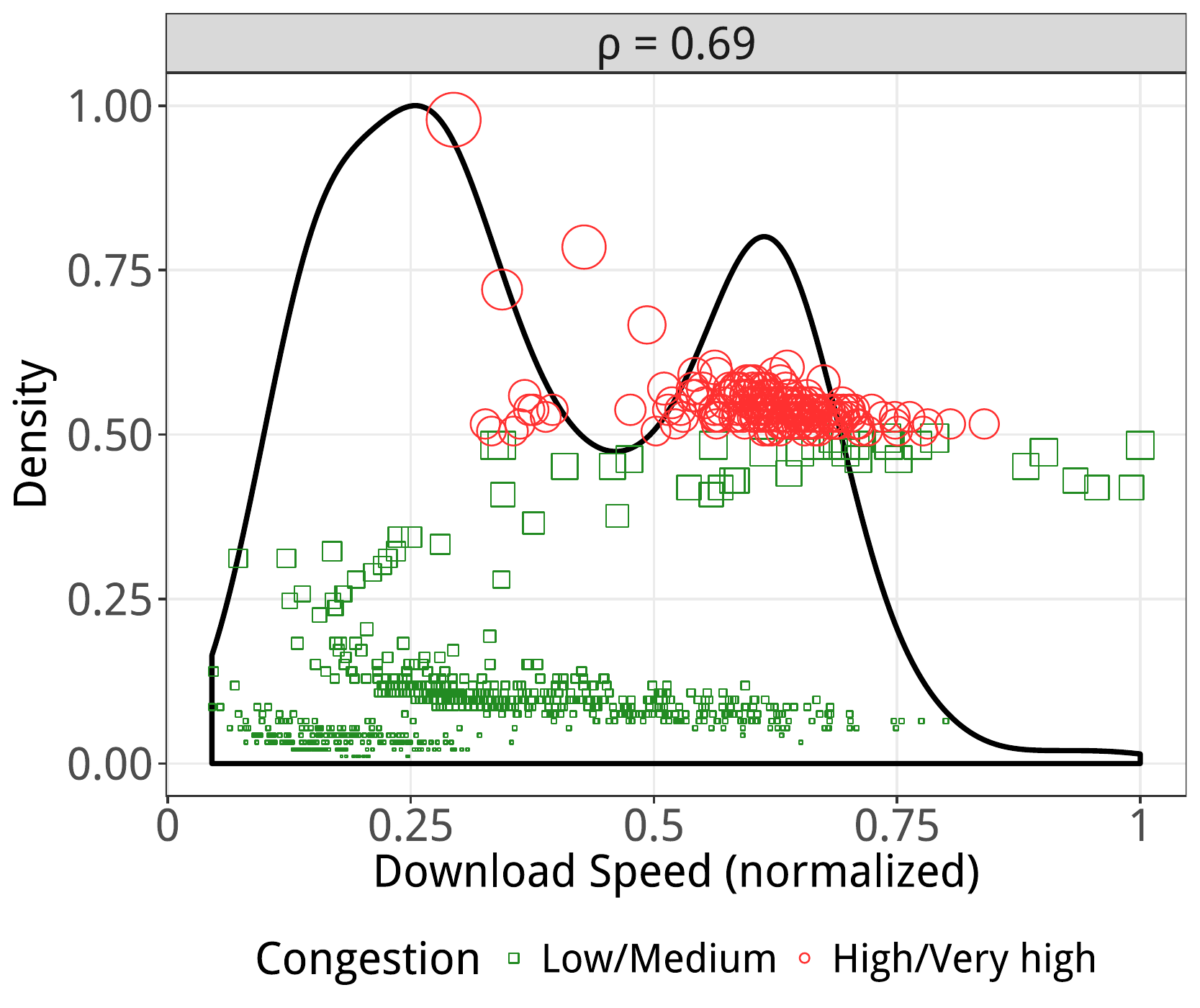}}\quad
				\label{fig:PosCorrAgg}
			}
		}
		\vspace{-5mm}
		\caption{Two samples of correlation between download-speed and congestion-count.}
		\vspace{-5mm}
		\label{fig:Corr}
	\end{center}
\end{figure*}

\section{Related Work and Background}\label{sec:prior}
%There is a wealth of prior work on measurement, estimation, and inference of broadband performance from various point of views. Researchers from MIT have debated the right ``measure'' to use \cite{MIT10}, and shown that for any testing methodology, attributing performance bottlenecks to constituent parts (network, broadband access link, ISP network, inter-connects) is technically challenging. Many years ago, some researchers had looked into estimating end-to-end available bandwidth analytically. In \cite{sigcomm02} a methodology and tool based on periodic packet streams is developed to estimate available bandwidth, and its relation to TCP throughput has been studied. The same authors in \cite{imc04} highlighted the various fallacies and pitfalls in bandwidth estimation, including the effects of multiple bottlenecks and cross-traffic. In \cite{sigmetrics07} a machine learning based model was develop to predict bandwidth availability. Yet, none of them touched real life measurement data to profile ISPs Residential broadband capacity.

While there is a commendable amount of effort being expended on collecting data, via either passive measurement of video traffic or active probing using hardware devices (we refer the reader to a recent survey \cite{survey15} that gives an overview of measurement platforms and standardization efforts), less effort has been expended on a systematic analysis of the collected data. This matters, because early works such as \cite{MIT10} have demonstrated that broadband speed measurements can exhibit high variability, and these differences arise from a complex set of factors including test methodology and test conditions, including home networks and end users' computers, that make it very challenging to attribute performance bottlenecks to the constituent parts, specifically the ISP network. While their work acknowledges that broadband benchmarking needs to look beyond statistical averages to identify factors exogenous to the ISP, they do not offer any specific approaches for doing so. A separate body of work \cite{sigcomm02,imc04,sigmetrics07} explores model-driven and data-driven methods to estimate or predict end-to-end available bandwidth.
In \cite{sigcomm02} a methodology and tool based on periodic packet streams is developed to estimate available bandwidth, and its relation to TCP throughput has been studied. The same authors in \cite{imc04} highlighted the various fallacies and pitfalls in bandwidth estimation, including the effects of multiple bottlenecks and cross-traffic. In \cite{sigmetrics07} a machine learning based model was developed to predict bandwidth availability. However, they operate at short time-scale, their data-sets are small, and their focus is not specific to broadband networks.

%We believe our work is among the first to combine causal inference techniques for observational studies with the big data openly available from the M-Lab measurement platform to attempt a fair comparison of ISP broadband performance. 

%Over the last couple of years, multiple platforms have emerged for measuring and reporting consumer Internet performance, for which we refer the reader to this comprehensive survey \cite{survey15}.  Among them, 
The Measurement Labs (M-Lab) \cite{M-Lab} is a large scale measurement platform to which several tools have been attached. A measurement tool, called Network Diagnostic Test (NDT) \cite{NDT} is a sophisticated speed test tool that gathers millions of speed-test measurements from the globe. There are also other tools that are designed for diagnosing last-mile networks and end-user systems (NPAD \cite{NPAD}), or for characterizing the performance of home networks, such as, BISmark \cite{BISmark}. Most commercial speed test tools, such as Ookla \cite{Ookla}, do not publish their performance data and measurement method.
With the aim of enhancing Internet transparency, M-Lab makes all the measurement data publicly accessible via various means. It allows users to query structured data by using a SQL-like query tool called {\tt BigQuery} \cite{Bigquery}, as well as a means of downloading raw data from a cloud storage tool called {\tt Gsutil} \cite{Gsutil}. NDT data has gained increasing interests from research community because of its rich diagnostics data. In addition, the volume of data collected by M-Lab, of the order of hundreds of millions of tests per year, makes NDT a precious big data source for studies to statistically draw meaningful inference from measurement data. There are some studies  using NDT data to measure latency variation in the Internet, and its results have been published in \cite{NDT_RTT}. 
We believe our work is among the first to estimate broadband capacity with the big data openly available from the M-Lab measurement platform to attempt a fair inference of ISP broadband performance.

\begin{figure*}[!t]
	\begin{center}
		\mbox{
			\subfigure[Single household 1.]{
				{\includegraphics[width=0.30\textwidth,height=0.20\textheight]{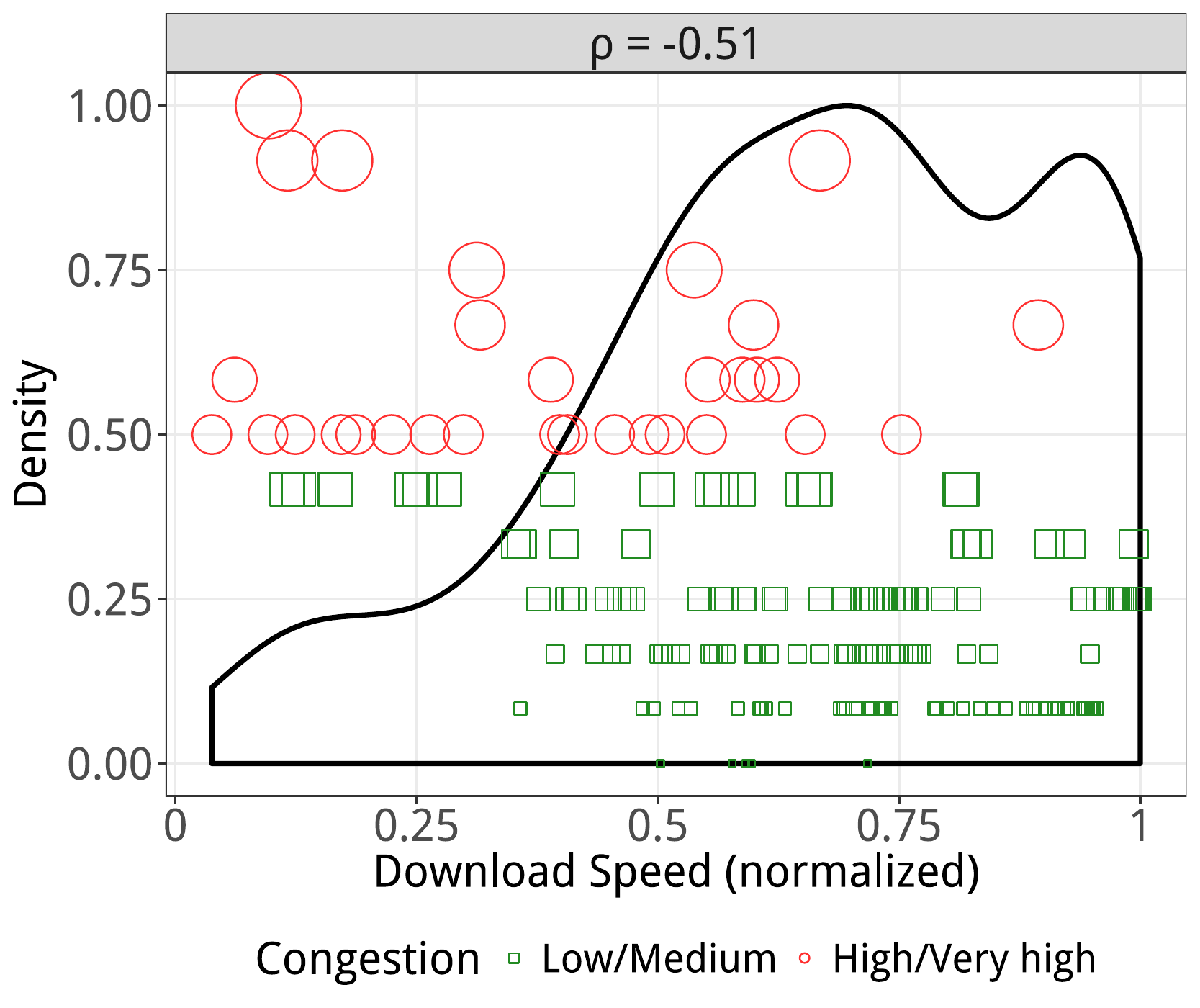}}\quad
				\label{fig:GThassan}
			}
			\hspace{0mm}
			\subfigure[Single household 2.]{
				{\includegraphics[width=0.30\textwidth,height=0.20\textheight]{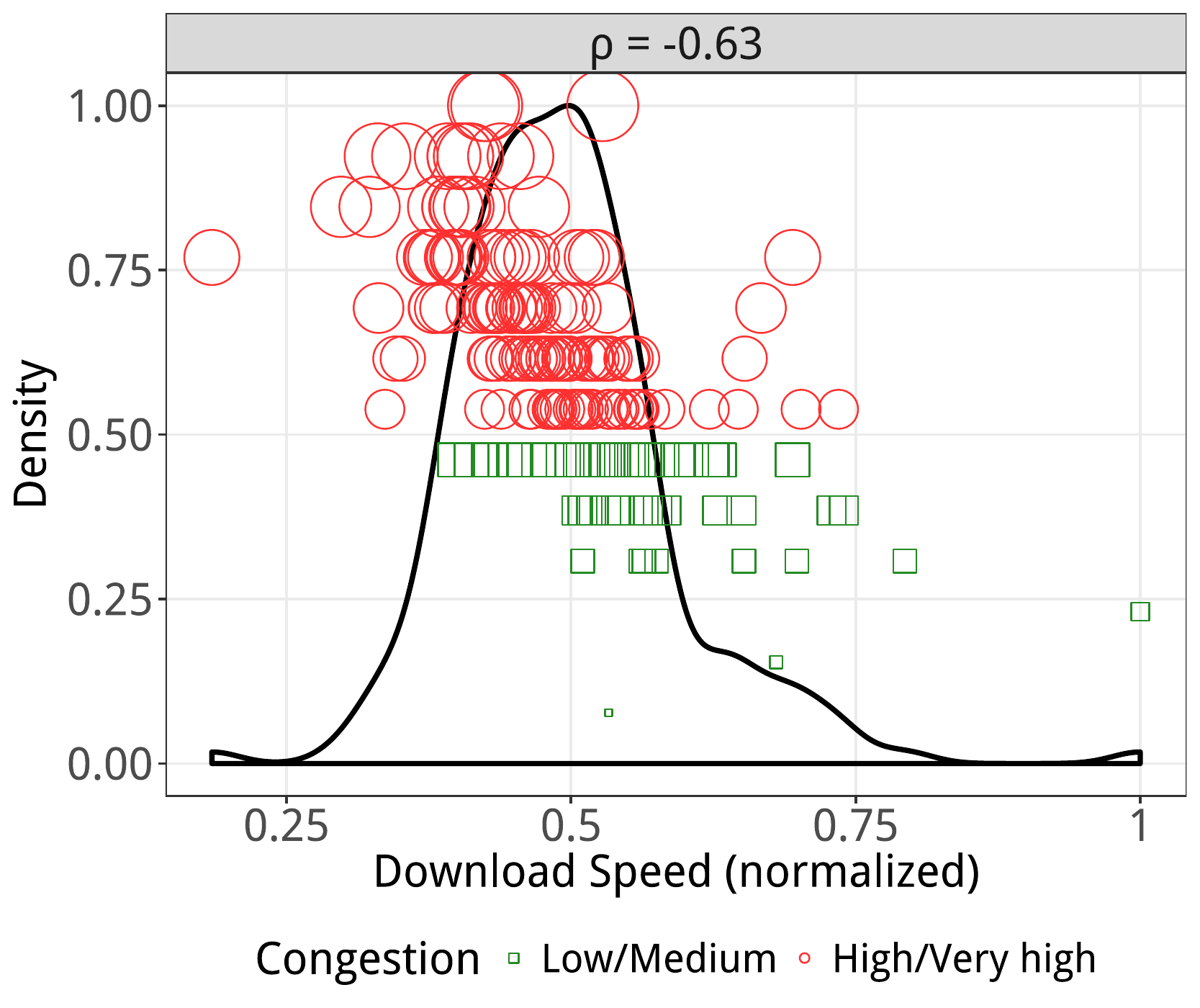}}\quad
				\label{fig:GTyun}
			}
			\hspace{0mm}
			\subfigure[Multiple households 1 \& 2.]{
				{\includegraphics[width=0.30\textwidth,height=0.20\textheight]{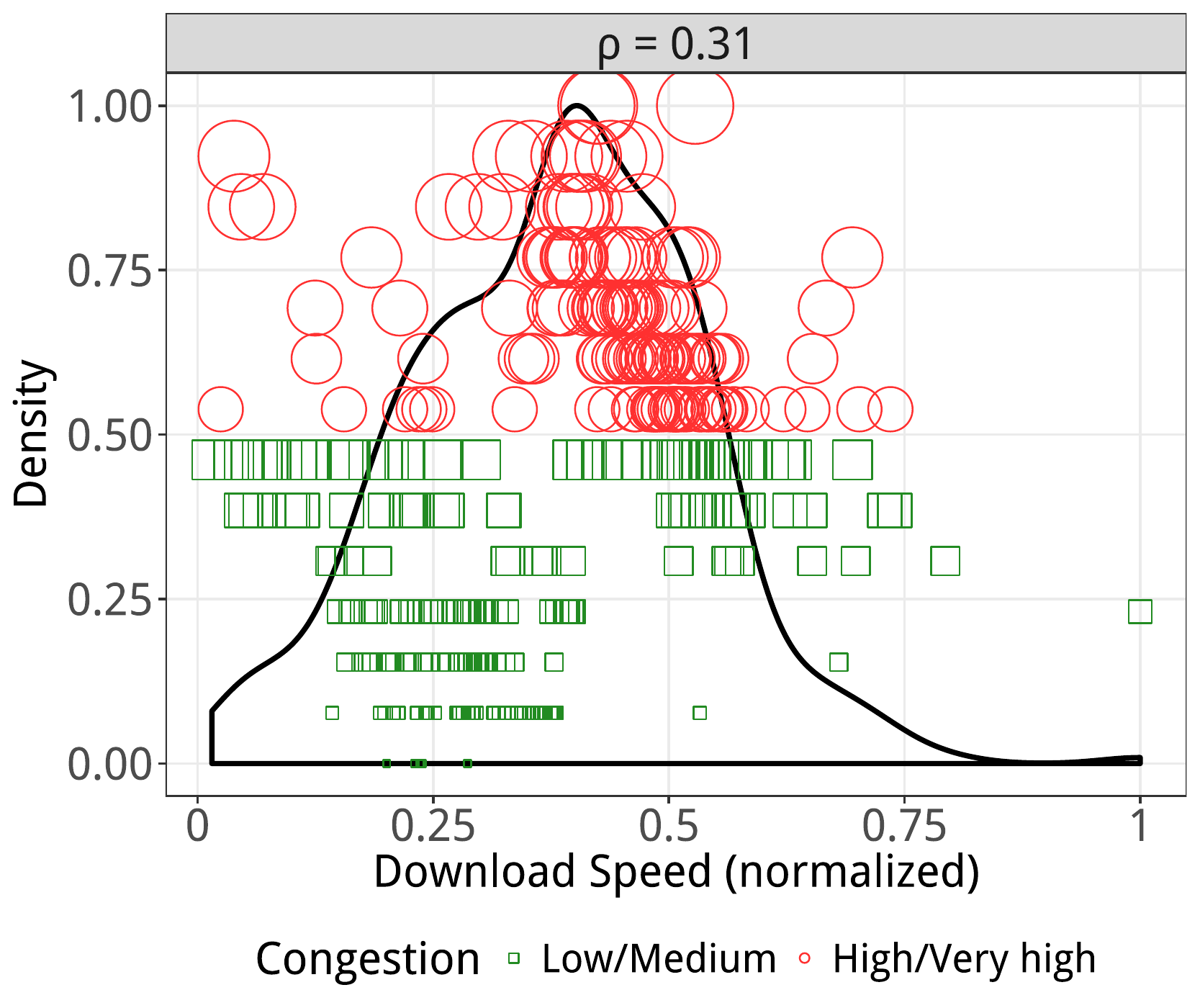}}\quad
				\label{fig:GThassanyun}
			}
		}
		\vspace{-5mm}
		\caption{Our ground truth measurements showing the correlation between speed and congestion count from: (a,b) each of known single houses, and (c) houses 1 \& 2 combined.}
		\vspace{-5mm}
		\label{fig:GT}
	\end{center}
\end{figure*}

%\subsection{NDT Dataset}
The NDT tool runs a throughput measurement in each direction: from client to server (upstream), and server to client (downstream), as follows. A client initiates an NDT measurement to M-Lab, and the M-Lab backend selects a server closest to the client using IP geolocations. Each test estimates the downstream throughput from the server to the client. The server logs statistics including round trip time, bytes sent, received, and acknowledged, congestion window size (CWND), and the number of congestion signals (multiplicative downward congestion window adjustments) received by the TCP sender. M-Lab makes the server logs statistics data available which makes it research friendly. The data are publicly available through Google’s BigQuery and Cloud Storage [29, 30]. Unlike Ookla and DSLreports, which only make aggregated stats available to public.

In this paper we use the data collected by the Network Diagnostic Test (NDT) tool, because it has by far the largest number of speed test samples (over 40 million for the year 2016), and captures a rich set of TCP statistics for each test. In order to evaluate the generality of our methods, we apply them to data from two countries: Australia (AU) and the United States (US). We select two large and two small ISPs from each country for comparison: Telstra, Optus, CEnet, and Harbour from AU; and Comcast, AT\&T, Lightower and Hurricane from the US. For these ISPs, we analyze the NDT speed test measurements taken over the twelve months, from 1 August 2016 to 31 July 2017, comprising $2.8$m samples for AU, and $17.4$m for US -- the latter is an order of magnitude larger since Google searches in the US got linked with NDT as of July 2016.

\section{Isolating Households}\label{sec:isolation}

%challenging, indexed by IP address
M-Lab data-points are indexed by IP address of home gateways. ISPs allocate IP addresses based on their resource pool, subscriber base, or business policy. In some cases, an ISP (often a large one) may have a fairly large pool of public IP addresses and can assign every subscriber a unique public IP, but one-to-one address lease may change dynamically over time. In other cases, the ISP (often a small one) will instead assign a public IP address to a group of subscribers, and then employ NAT to multiplex their traffic. Consequently, it becomes challenging to extract the broadband capacity from M-Lab data, as an IP address does not necessarily represent a single  household. Thus, we need a method to isolate data-points corresponding to single households.   

%congestion vs. speed
The congestion signal of each NDT data-point indicates that how TCP congestion window (\textit{cwnd}) is being affected by congestion, and is incremented by any form of congestion notifications including Fast Retransmit, ECN, and timeouts. Intuitively, a large value of congestion signal (congestion count) should correspond to a low TCP throughput (download speed), and vice versa. 

We denote the \textbf{Pearson's correlation coefficient} between the measured download speed and recorded congestion count by $\rho$. This parameter is computed across all tests corresponding to a given client IP address.
We expect $\rho$ to be negative for any given household, as higher broadband speed should correspond with lower congestion, and this is indeed the case for a majority of client IP addresses contained in the M-Lab data. However, for some IP addresses we observe strong positive correlations (i.e. $\rho>0$). Our hypothesis for this unexpected phenomenon is that when multiple houses of an ISP network are sharing an IP address; the speed measurements can vary in a wide range (e.g. [5, 50] Mbps) depending on broadband capacity of individual households, whereas congestion counts would have smaller variations (e.g. [6, 10]) reflecting the condition of the network. Thus, having mixed measurements (speed and congestion-count)  from multiple households will likely result in imbalanced data pairs causing an unexpected positive correlation between speed and congestion-count.

To better visualize our discussion and hypothesis, we present in Fig.~\ref{fig:Corr} samples of correlation between the download-speed and congestion-count observed over a 12-month period from two IP addresses. In each plot, the normalized density distribution of download-speed measurements is depicted by solid black lines. We overlay it by scatter plot of download-speed (x-axis) and its corresponding congestion-counts (y-axis), shown by square/circle markers. Note that for a given IP address, we unit-scale (normalize) measured download-speed and congestion-count separately by dividing each data point by corresponding maximum value (i.e. $X_i/X_{max}$ and $Y_i/Y_{max}$; where [$X_i$, $Y_i$] is the pair of download-speed and congestion-count for a client IP).  
In our plots, the scaled value of congestion count for each test-point is proportional to the size of corresponding marker, tiered in two colors -- low/medium (i.e. $<0.5$) congestion counts are in green, and high/very-high (i.e. $\geq0.5$)  congestion counts are in red. %The \textbf{\color{red}Pearson's} correlation coefficient (i.e. $\rho$) between download-speed and congestion-count is shown on top of each plot.  

Fig. \ref{fig:NegCorrAgg} shows a negative correlation (i.e. $\rho=-0.83$) for 458 test-points obtained from an IP address served by Cox ISP in the US -- smaller green squares are mainly skewed towards the bottom right of the plot (i.e. low congestion and high speed values), and larger red circles are grouped at the top left region of the plot (i.e. high congestion and low speed values). On the other hand, Fig. \ref{fig:PosCorrAgg} shows a positive correlation (i.e. $\rho=0.69$) for 896 test-points from City ISP in the US --  smaller green squares are mainly spread from left to middle bottom of the plot (i.e. low congestion and low/medium speed values), and larger red circles are clustered at top middle of the plot (i.e. high congestion and medium/high speed values). 

\textbf{Summary:} We believe that client IP addresses with negative $\rho$ values likely represent single households and are selected for our study. In other words, positive $\rho$ values indicate a plurality of houses (sharing addresses, or aggregated behind NAT) that their corresponding data points are filtered. 

In next section, we validate our hypothesis at small scale by comparing  to ground truth taken from two known houses, and at large scale by checking consistency within a network operator across months and across various network operators.

\begin{figure*}[t!]
	\begin{center}
		\mbox{
			\subfigure[Australia.]{
				{\includegraphics[width=0.495\textwidth,height=0.25\textheight]{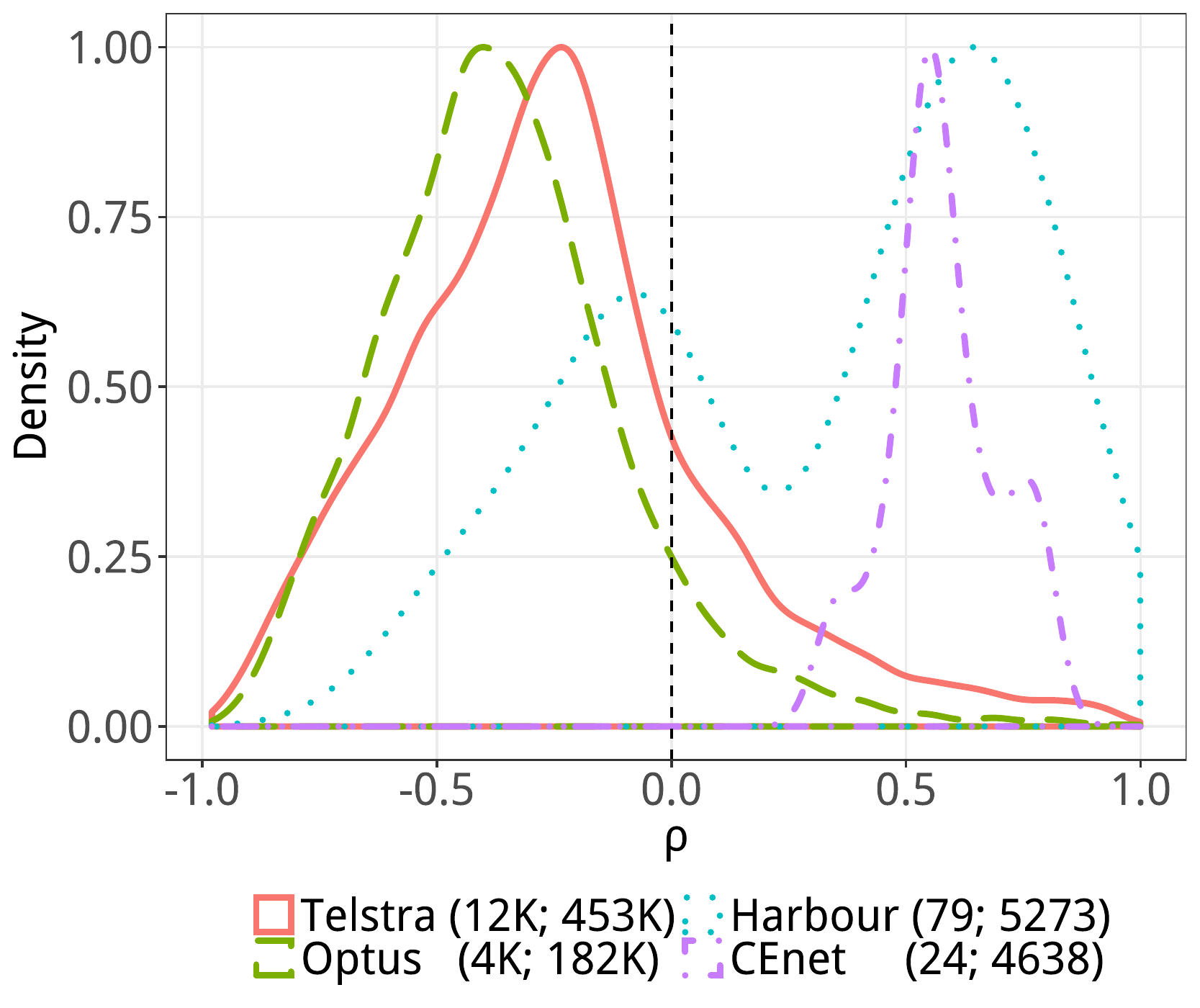}}\quad
				\label{fig:rhoAU}
			}
			\hspace{-2mm}
			\subfigure[The U.S.]{
				{\includegraphics[width=0.495\textwidth,height=0.25\textheight]{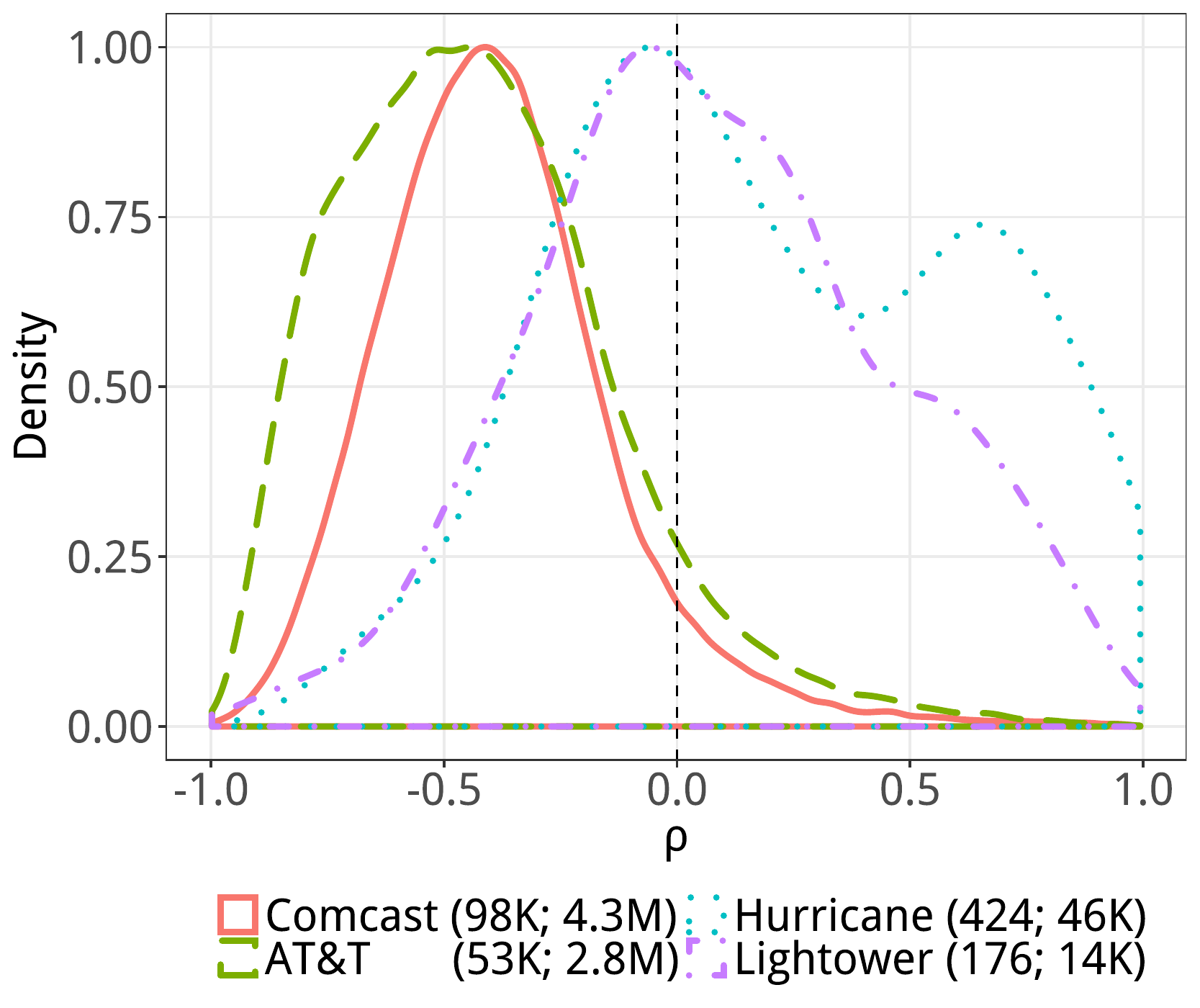}}\quad
				\label{fig:rhoUS}
			}
		}
		\vspace{-5mm}
		\caption{ Negative/Positive correlation across large/small ISPs in (a) AU, and (b) US.}
		\vspace{-5mm}
		\label{fig:rhoAUUS}
	\end{center}
\end{figure*}

%\begin{figure*}[t!]
%	\begin{center}
%		\mbox{
%			\subfigure[Australia.]{
%				{\includegraphics[width=0.495\textwidth,height=0.30\textwidth]{Fig/corrISP/AU_cor}}\quad
%				\label{fig:rhoAU}
%			}
%			\hspace{-2mm}
%			\subfigure[U.S.]{
%				{\includegraphics[width=0.495\textwidth,height=0.30\textwidth]{Fig/corrISP/US_cor}}\quad
%				\label{fig:rhoUS}
%			}
%		}
%		\vspace{-5mm}
%		\caption{ Negative/Positive correlation across large/small ISPs in (a) Australia, and (b) U.S.}
%		\vspace{-5mm}
%		\label{fig:rhoAUUS}
%	\end{center}
%\end{figure*}

\section{Validation of Our Approach}\label{sec:validation}
We now inspect the validity of our hypothesis whether the value of $\rho$ is the key parameter for identifying single household data points.

\subsection{Ground Truth Validation at Small Scale}
We first examine our hypothesis empirically by conducting performance test using NDT client program in two known households (of authors of this paper). Our test houses have different capacities of broadband link -- 8 Mbps for household-1 and 20 Mbps for household-2. We wrote a Python script that periodically runs the standard NDT client. We collected 200 test points in each house over two weeks in December 2017. We now compute the $\rho$ parameter and visualize data of individual households as well as combined data of the two houses, as shown in Fig.~\ref{fig:GT}.  

It is seen that the $\rho$ value for each single house individually is negative, i.e.  $\rho=-0.51$ for household-1 in Fig.~\ref{fig:GThassan} and $\rho=-0.63$ for household-2 in Fig.~\ref{fig:GTyun}, with red circles mainly clustered to the left of the plot (high congestion and low speed) and green squares skewed to the right region (fairly low congestion and medium/high speed). However, combining data of the two houses causes a cluster of green squares to emerge on the lower left of the plot (low congestion and low speed) and red circles to slightly shift towards the right side (high congestion and high speed). This results in a positive correlation between download-speed and congestion-count, i.e.  $\rho=0.31$ in Fig.~\ref{fig:GThassanyun}.

Overall, results of this experiment validate our hypothesis (in \S\ref{sec:isolation}) at small scale with ground truth data of known households.

\begin{figure*}[t!]
	\begin{center}
		\mbox{
			\subfigure[Consistency of negative correlation (Cox, 458 tests from 98.174.39.22).]{
				{\includegraphics[width=0.495\textwidth,height=0.23\textheight]{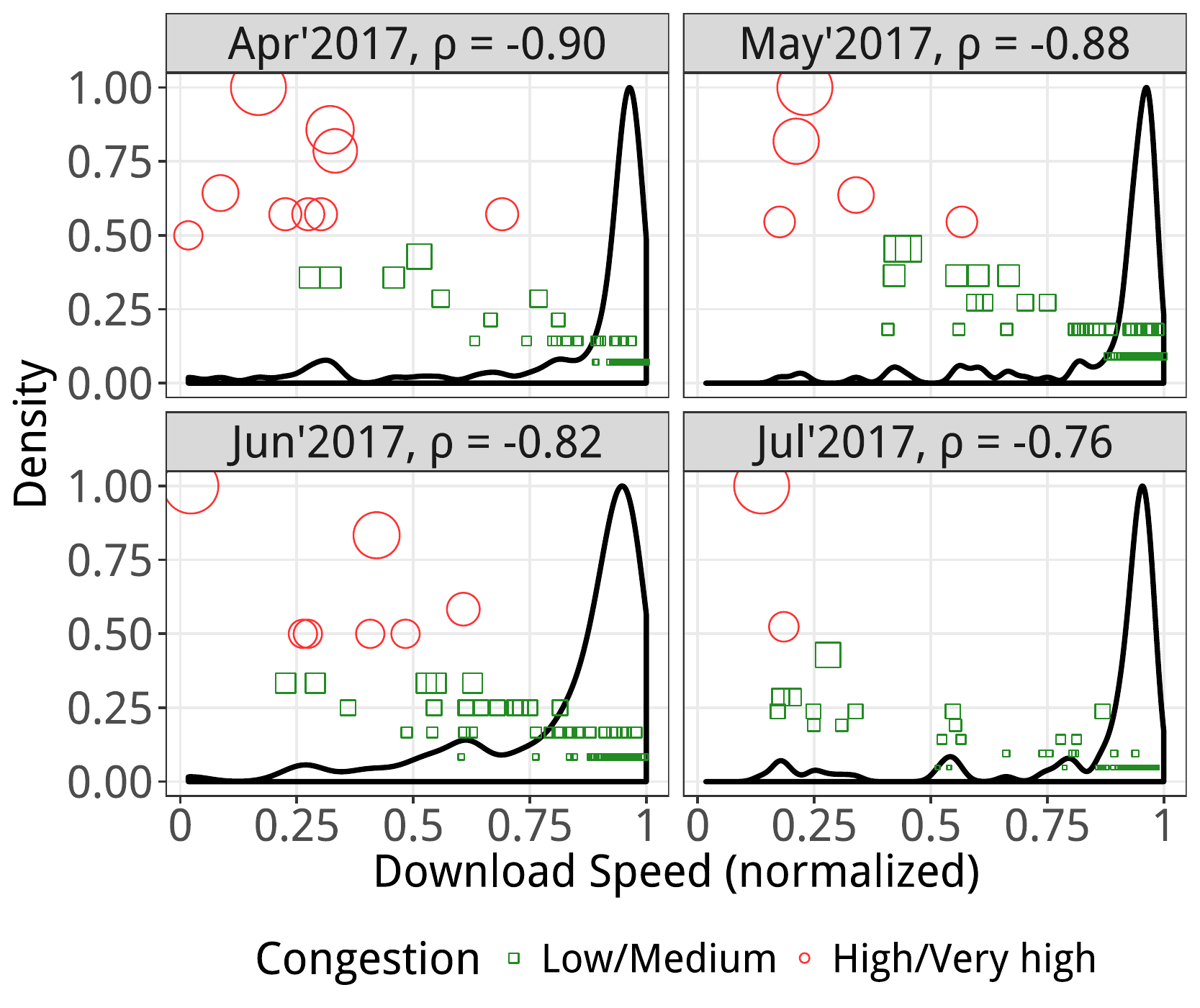}}\quad
				\label{fig:NegCorrMnt}
			}
			\hspace{-2mm}
			\subfigure[Consistency of positive correlation (City, 896 tests from 64.39.155.194).]{
				{\includegraphics[width=0.495\textwidth,height=0.23\textheight]{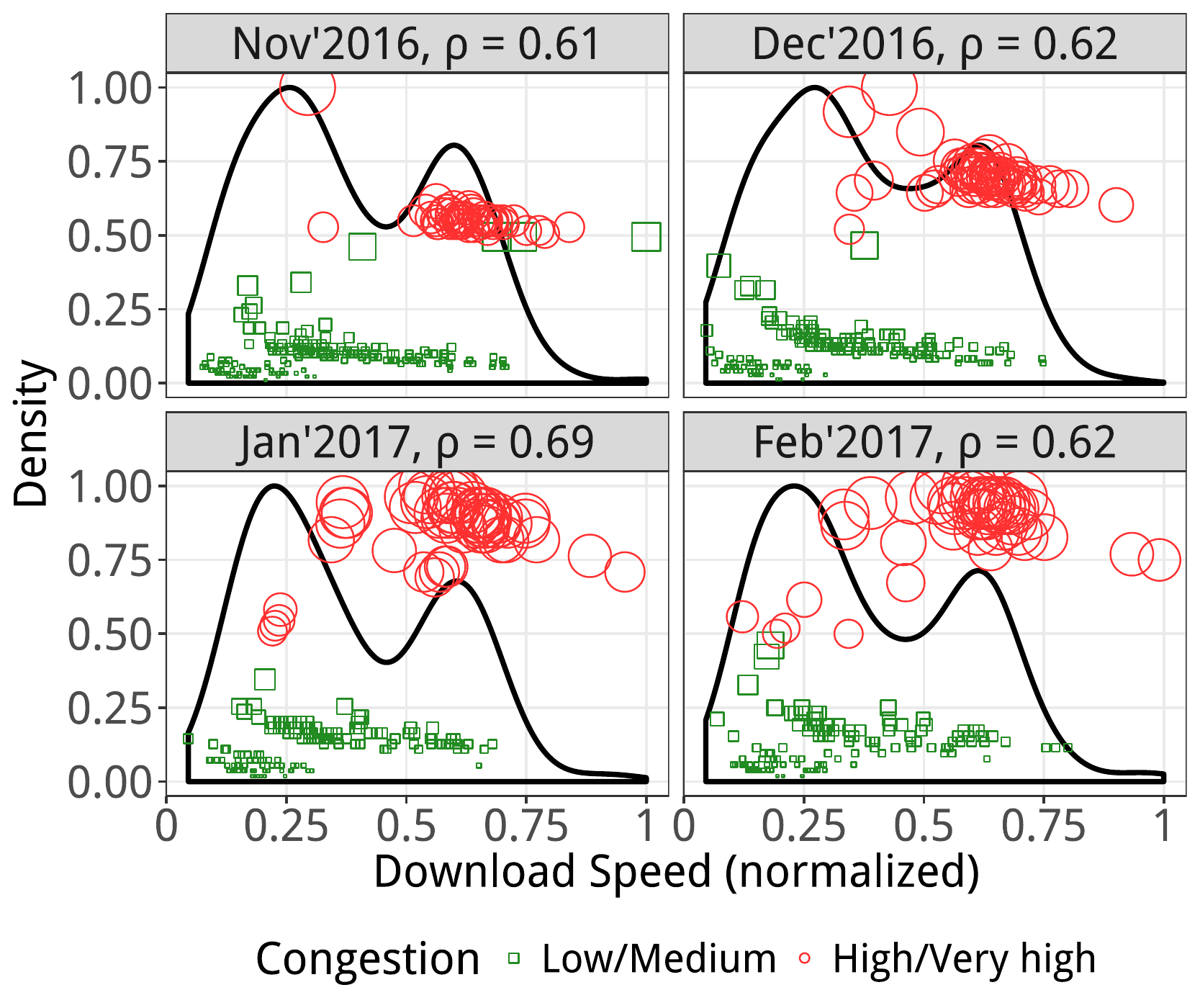}}\quad
				\label{fig:PosCorrMnt}
			}
		}
		\vspace{-5mm}
		\caption{Consistency of correlation between download-speed and congestion-count across four months: (a) negative, and (b) positive correlation.}
		\vspace{-5mm}
		\label{fig:CorrMnt}
	\end{center}
\end{figure*}

\begin{figure*}[t!]
	\begin{center}
		\mbox{
			\subfigure[Australia.]{
				{\includegraphics[width=0.495\textwidth,height=0.30\textwidth]{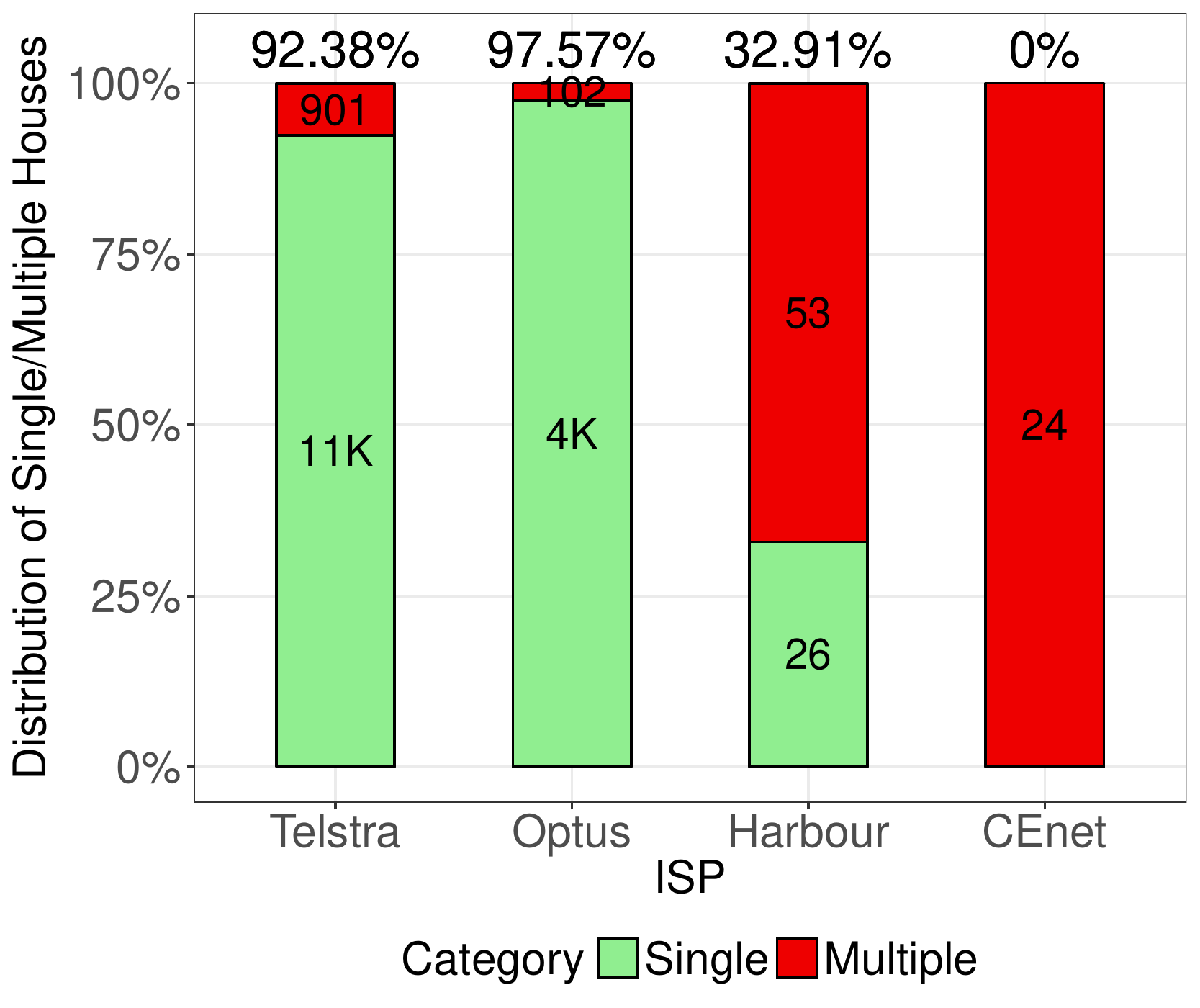}}\quad
				\label{fig:FilterSmapleCountAU}
			}
			\hspace{-2mm}
			\subfigure[The U.S.]{
				{\includegraphics[width=0.495\textwidth,height=0.30\textwidth]{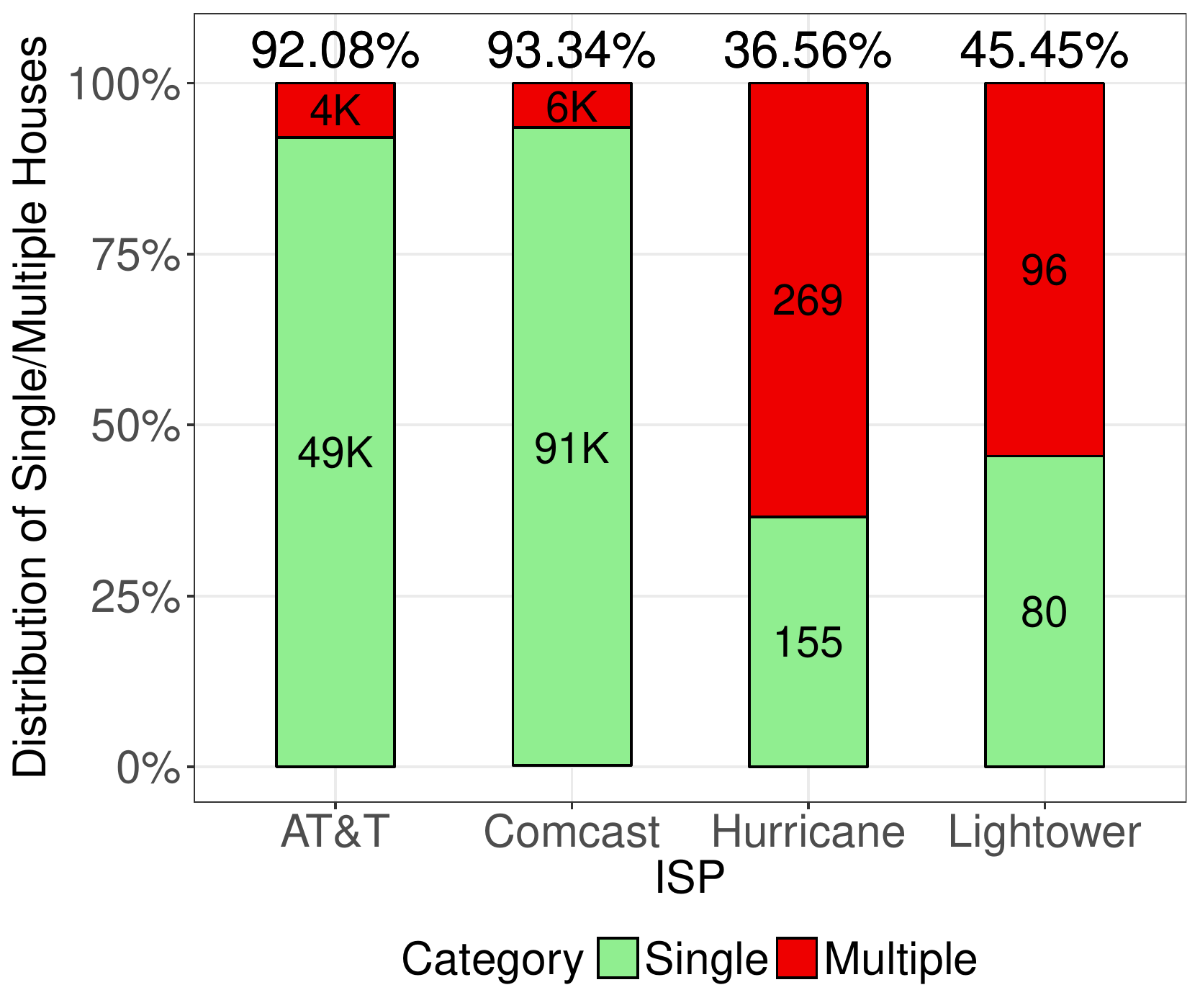}}\quad
				\label{fig:FilterSmapleCountUS}
			}
		}
		\vspace{-5mm}
		\caption{Filtering based on annual sample count of each unique IP address across large/small ISPs in (a) AU, and (b) US.}
		\vspace{-5mm}
		\label{fig:FilterSmapleCount}
	\end{center}
\end{figure*}

\begin{figure*}[t!]
	\begin{center}
		\mbox{
			\subfigure[Many outliers in speed measurements.]{
				{\includegraphics[width=0.30\textwidth,height=0.20\textheight]{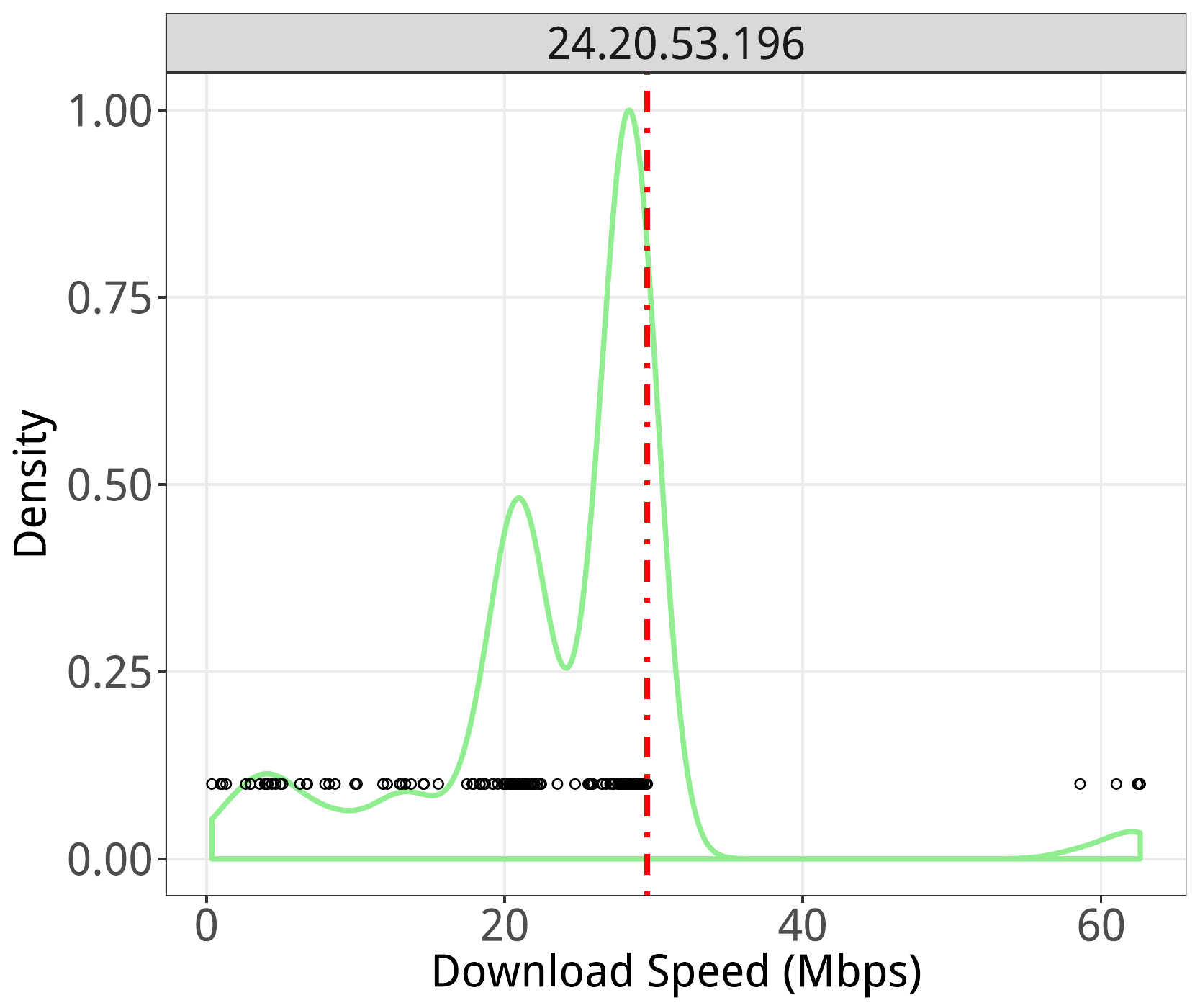}}\quad
				\label{fig:outliersMany}
			}
			\hspace{0mm}
			\subfigure[One outlier in speed measurements.]{
				{\includegraphics[width=0.30\textwidth,height=0.20\textheight]{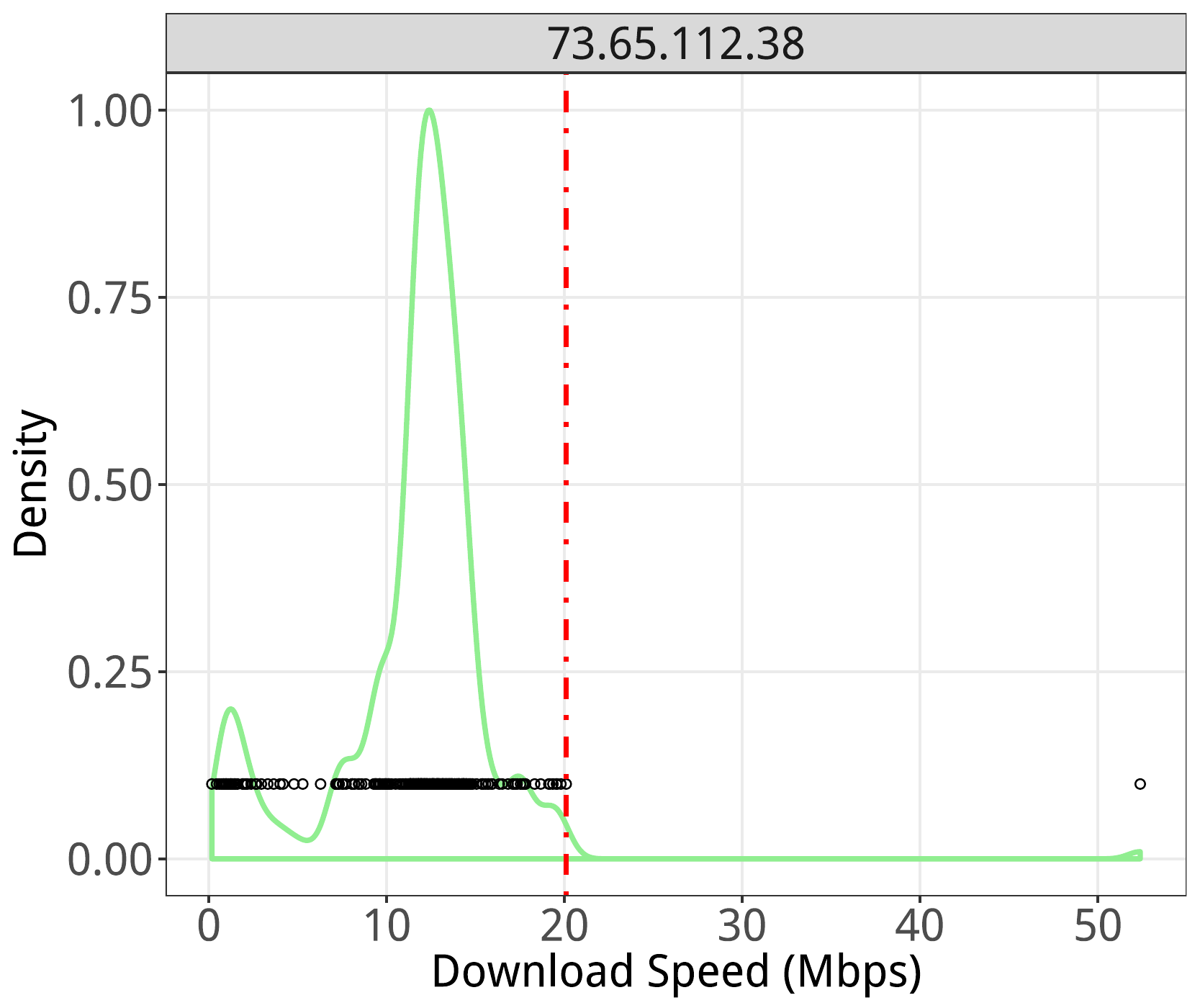}}\quad
				\label{fig:outliersOne}
			}
			\hspace{0mm}
			\subfigure[No outlier in speed measurements.]{
				{\includegraphics[width=0.30\textwidth,height=0.20\textheight]{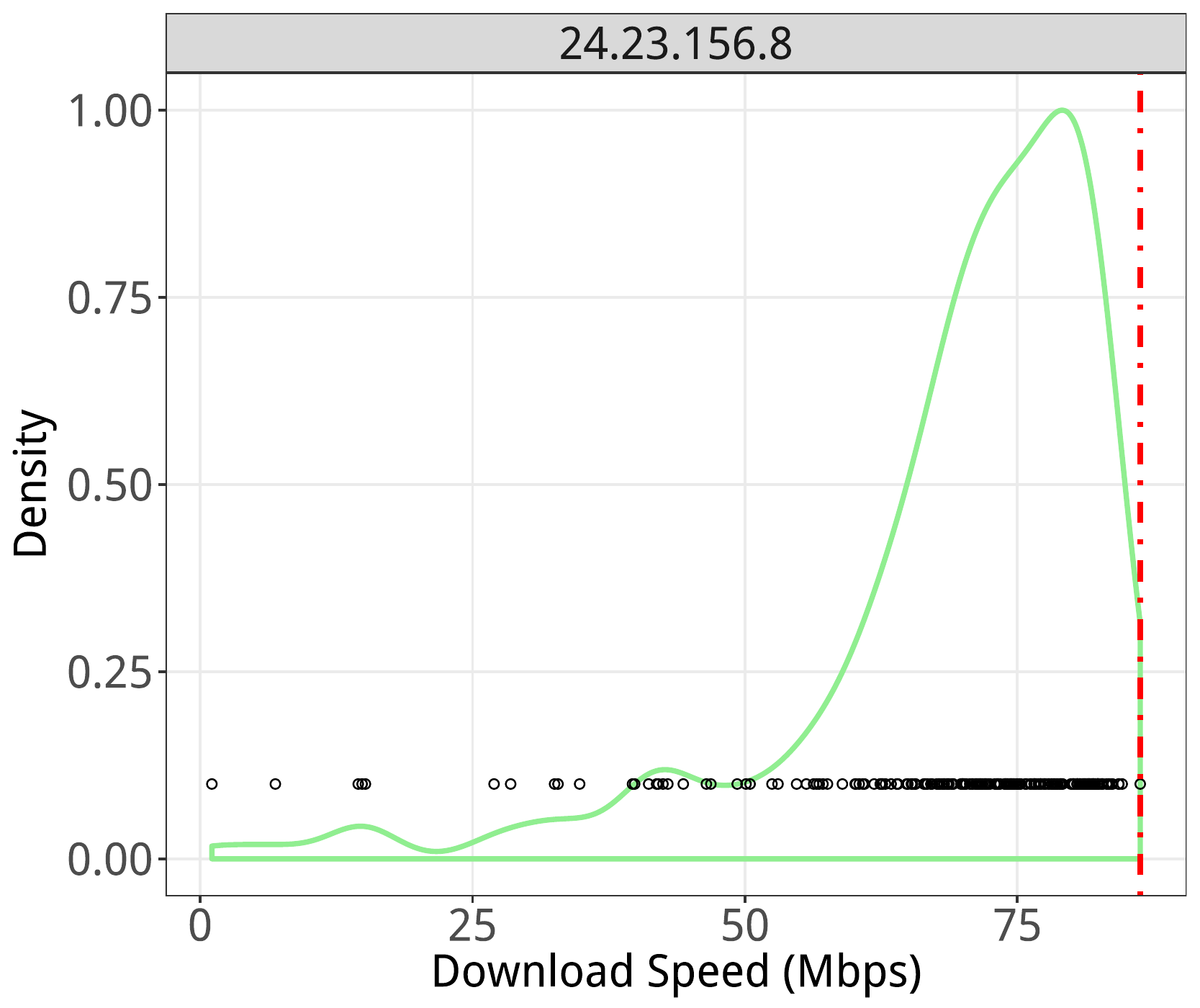}}\quad
				\label{fig:outliersZero}
			}
		}
		\vspace{-5mm}
		\caption{Outliers in speed measurements.}
		\vspace{-6mm}
		\label{fig:outliers}
	\end{center}
\end{figure*}

\begin{table}[!h]
	\centering
	\caption{Number of public IPv4 addresses and subscriber counts for selected ISPs.}
	\label{table:numIP}
	\vspace{-0.2cm}
	\begin{tabular}{|l|c|r|}
		\hline 
		\textbf{ISP} & \textbf{Num. of public addresses} & \textbf{Subscriber counts} \tabularnewline
		\hline 
		AT\&T (US)  &  91,230,208 & 14.2m \cite{ATTSubs}\tabularnewline
		\hline
		Comcast  (US)  & 50,872,320 & 23.3m \cite{ComcastSubs}  \tabularnewline
		\hline
		Hurricane (US) & 356,352  & N/A          \tabularnewline
		\hline
		Lightower (US) & 124,160   & 1m \cite{LightowerSubs}       \tabularnewline
		\hline
		Optus (AU)  	 & 5,230,337   & 1.2m \cite{OptusSubs}   \tabularnewline
		\hline
		Telstra (AU) 	 & 13,115,648  & 3.5m \cite{TelstraSubs}   \tabularnewline
		\hline
		Harbour (AU)	  & 2,816       & at least 5K \cite{HarbourSubs} \tabularnewline
		\hline
		CEnet (AU) 	 & 4,096     &  330K \cite{CenetSubs}		 \tabularnewline                       
		\hline 
	\end{tabular}
	\vspace{-5mm}
\end{table}

\subsection{Consistency Validation at Large Scale}
We now go back to our M-Lab data to analyze the $\rho$ parameter across various ISPs of different size as well as across months checking whether a consistent pattern of correlation is observed.

\subsubsection{Across ISPs}
Large ISPs such as AT\&T and Comcsat in the US with a wealth of public IP addresses (i.e. 91 million and 51 million) 
Smaller ISPs who own smaller pool of IPv4 addresses (e.g. class C blocks) are more likely forced to employ NAT (or dynamic lease in the best case) for better management of their limited address resources. On the other hand, larger ISPs who were assigned class A address blocks would have discretion to statically allocate one public IP address to each of their clients.   

Table~\ref{table:numIP} shows the total number of IPv4 public addresses along with the number of subscribers for selected ISPs in AU and US. The subscriber counts reported online may not be that accurate or up-to-date but it is apparent that all large ISPs namely AT\&T, Comcast, Telstra, and Optus, have far more number of public IPv4 addresses than their total number of subscribers. On the other hand, we see that subscriber counts is significantly larger (sometimes one order of magnitude) than the total number of IP addresses owned by small ISPs including Lightower, Harbour, and CEnet. % r Smaller ISPs in both AU and US, on the other hand, own Lightower has a an approximate subscriber number of 1,048,861 \cite{LightowerSubs} which is nearly ten folds of its total number of public IPv4 address. Similarly, Harbour has less than 3K public IPv4, yet as per \cite{HarbourSubs} it has more than 5K subscribers, and CEnet has only 4K public IPv4 yet approximately 330K users.

We, therefore, start examining the aggregate $\rho$ parameter for each ISP with focus on two countries namely Australia and US. We select two of large and two of small ISPs from each country for comparison: in Australia, Telstra and Optus as large providers, and Harbour and CEnet as small providers; in the US, Comcast and AT\&T as large providers, and Hurricane and Lightower as small providers. We present in Fig.~\ref{fig:rhoAUUS} the normalized density distribution of $\rho$ value across unique IP addresses of each ISP.
We find $\smallsim$12K, $\smallsim$4K, 79, and 24 unique addresses from network of Australian ISPs Telstra, Optus, Harbour and CEnet respectively conducting total of $\smallsim$453K, $\smallsim$182K, 5273, and 4638 NDT tests over 12-month period (Aug'16 - Jul'17). Fig.~\ref{fig:rhoAU} shows the $\rho$ distribution for our selected operators in Australia. It is seen that the $\rho$ parameter is predominately negative in large ISPs (shown by solid red lines for Telstra and dashed green lines for Optus in Fig.~\ref{fig:rhoAU}) suggesting that majority of IP addresses present in M-Lab data (from these two large ISPs) are consistently assigned to single households. Moreover, the $\rho$ distribution is fairly biased towards positive values for smaller ISPs -- average $\rho= 0.31$ for Harbour (its distribution is shown by dotted blue lines) and average $\rho= 0.58$ for CEnet (its distribution is shown by dashed-dotted purple lines) in Fig.~\ref{fig:rhoAU}, meaning that IP addresses are mainly shared by multiple households of varied broadband capacity.       

Similarly, we observe an aggregate negative correlation values for large ISPs in the US along with neutral/positive correlation for smaller ISPs, as shown in Fig.~\ref{fig:rhoUS}. For our US selected ISPs Comcast, AT\&T, Hurricane, and Lightower, we have $\smallsim$4.3m, $\smallsim$2.8m, $\smallsim$46K, and $\smallsim$14K NDT test-points respectively indexed by $\smallsim$98K, $\smallsim$53K, 424, and 176 unique addresses. The average $\rho$ for large operators Comcast and AT\&T is  $-0.39$ and $-0.43$ respectively, whereas smaller operators Hurricane and Lightower exhibit positive average correlation of $0.21$ and $0.10$ respectively. 

We see on average a negative correlation between measured download-speed and congestion-count across large network operators (with large pool of IP addresses), and positive correlation values across small network operators (with small pool of IP addresses) in two countries Australia and US.

\subsubsection{Across Months}
We now track the correlation value within a network operator across various months to check whether change of network conditions would affect the $\rho$ value. This verifies the validity of our hypothesis over time. We, therefore, compute the $\rho$ value for a given IP address on a monthly basis using data points observed within a month, e.g. April 2017. 

In Fig.~\ref{fig:Corr}, we saw speed, congestion and the $\rho$ value computed on aggregate data of 12-month period for one sample of IP address in each network (large and small separately). We now visualize in Fig.~\ref{fig:CorrMnt}  the monthly data along with corresponding $\rho$ values for the same IP addresses and their respective networks. 

We consistently observe an strong negative correlation for data of address \textit{98.174.39.22} from Cox (one of the top ten large ISPs in the US) across four months in 2017, as shown in Fig.~\ref{fig:NegCorrMnt}. Individual monthly speed density curves (narrow single hump) and congestion clusters are fairly similar to the plot in Fig.~\ref{fig:NegCorrAgg}, and the $\rho$ value is $-0.90$, $-0.88$, $-0.82$, and $-0.76$ for successive months April, May, June, and July respectively. 

Considering the IP address from a smaller operator in Fig.~\ref{fig:PosCorrMnt}, a strong positive correlation is observed consentingly across four successive months in 2016-2017. In each plot, download-speed density curve depicts two humps and congestion markers are aligned (green squares on the left and red circles on the right) in opposite direction to that which is expected, just similar to aggregate performance measurements in Fig.~\ref{fig:PosCorrAgg}. We again see a strong positive $\rho$ values of $0.61$, $0.62$, $0.69$, and $0.62$ respectively for November and December in 2016, and January and February in 2017. 

Our analysis of M-Lab data across various network operators and across various months validates that our hypothesis holds true at large scale too.

\begin{figure}[t!]
	\centering
	\includegraphics[width=0.475\textwidth,height=0.30\textwidth]{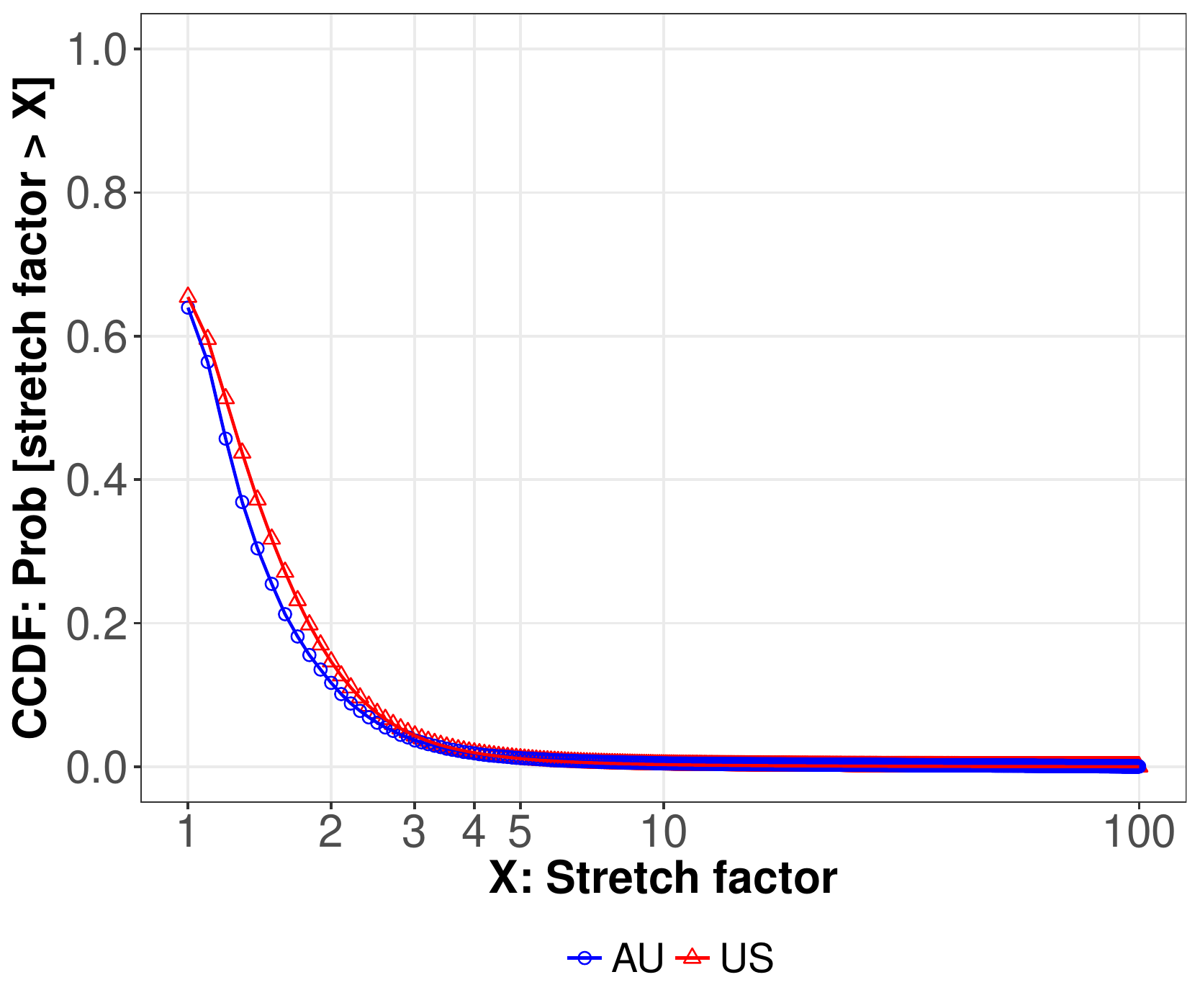}
	\vspace{-5mm}
	\caption{CCDF of stretch factor.}
	\label{fig:ccdf}
	\vspace{-5mm}
\end{figure}

\begin{figure*}[t!]
	\begin{center}
		\mbox{
			\subfigure[AT\&T (raw data -- each IP treated as a house).]{
				{\includegraphics[width=0.495\textwidth,height=0.30\textwidth]{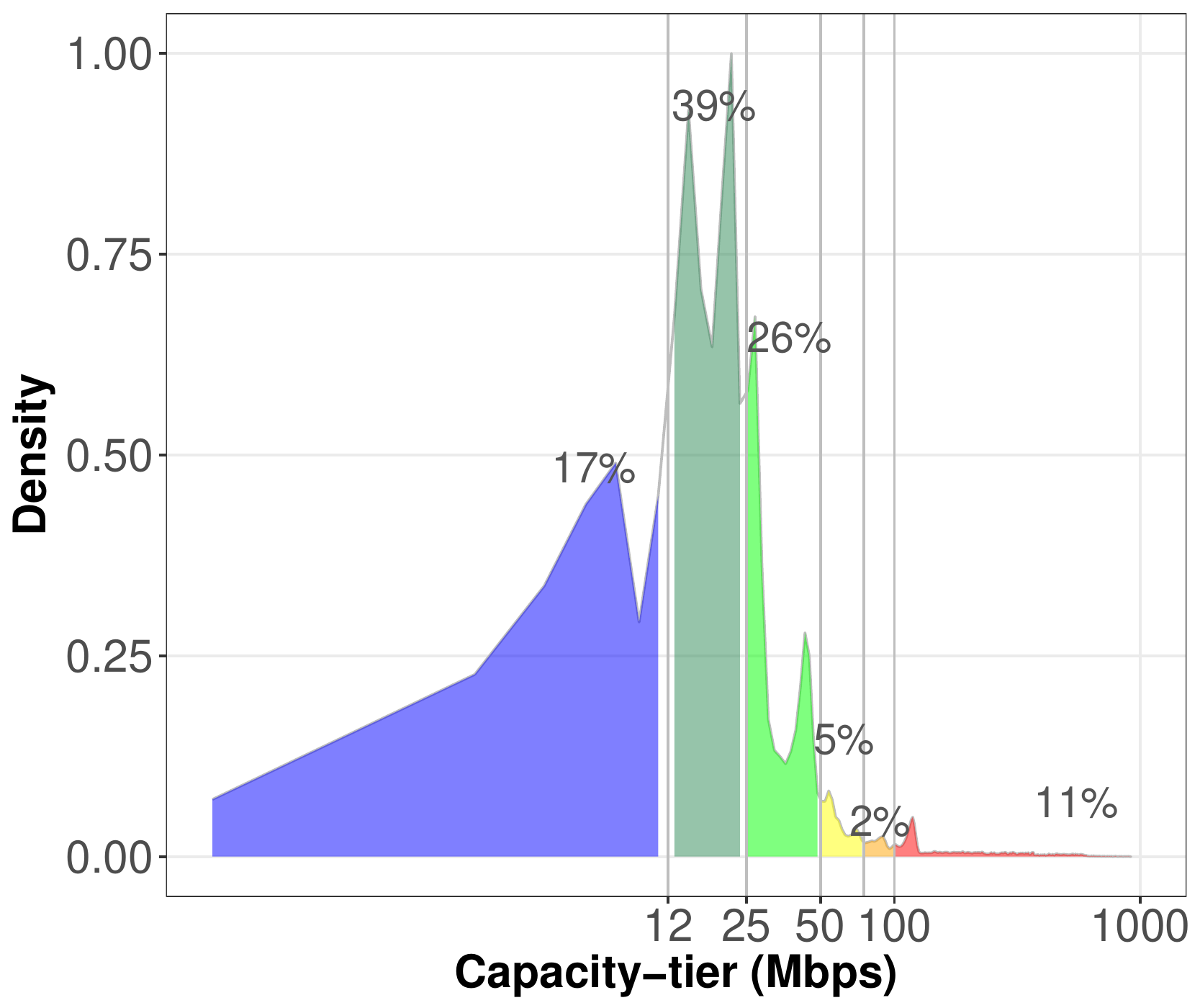}}\quad
				\label{fig:ATTraw}
			}
			\hspace{-2mm}
			\subfigure[Hurricane (raw data -- each IP treated as a house).]{
				{\includegraphics[width=0.495\textwidth,height=0.30\textwidth]{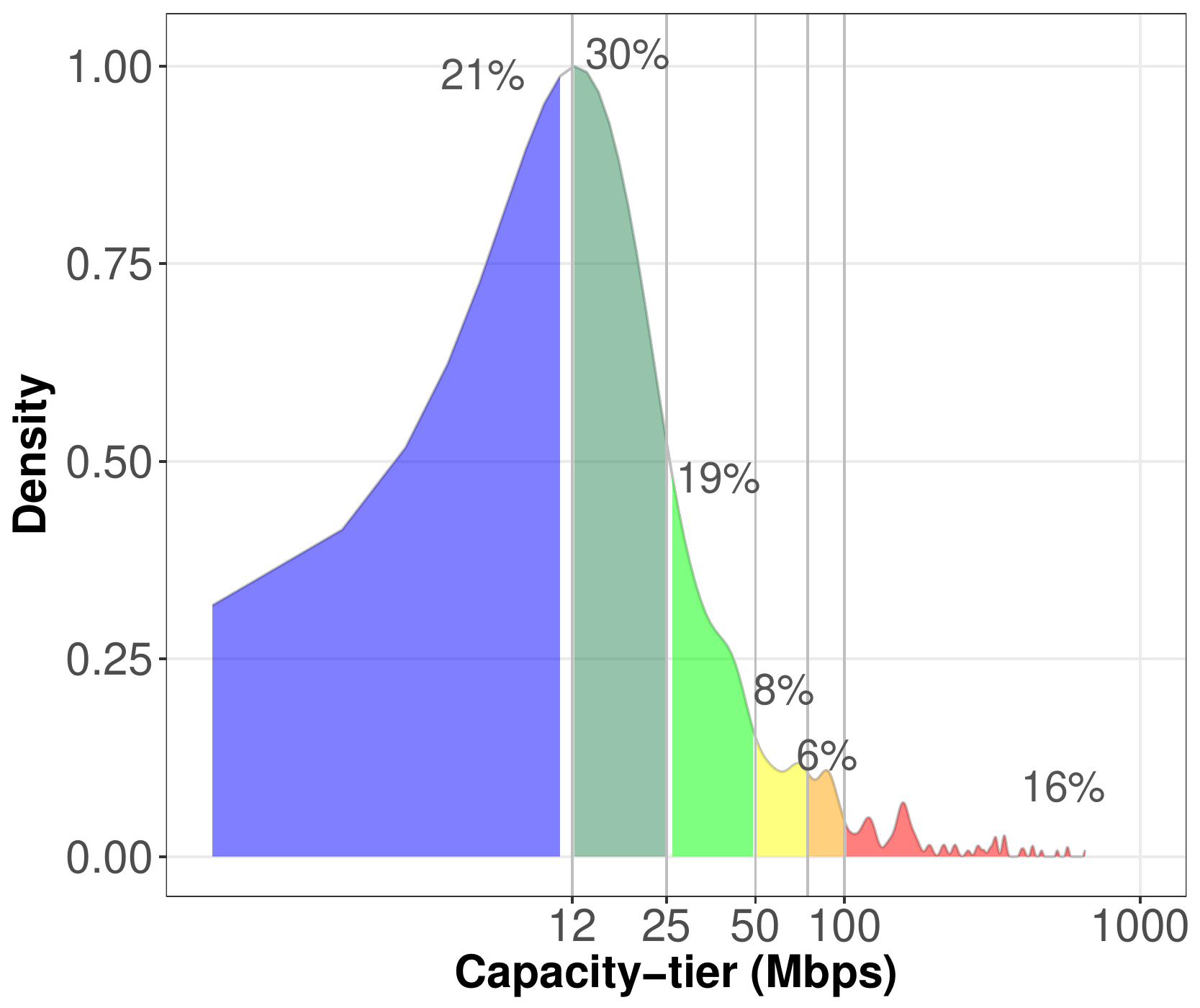}}\quad
				\label{fig:Hurricaneraw}
			}
		}

		\mbox{
			\subfigure[AT\&T (after removing houses with positive $\rho$).]{
				{\includegraphics[width=0.495\textwidth,height=0.30\textwidth]{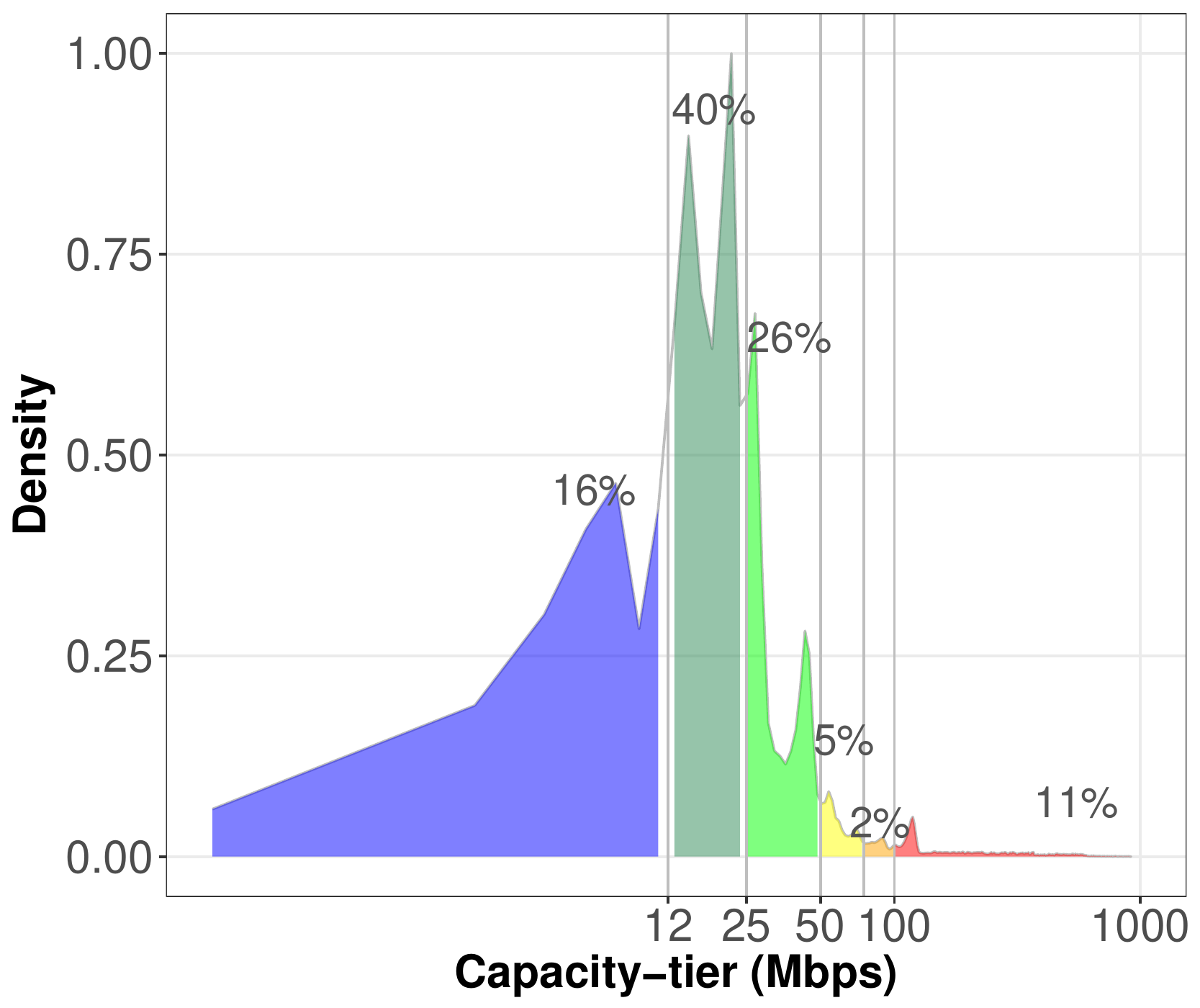}}\quad
				\label{fig:ATTbefore}
			}
			\hspace{-2mm}
			\subfigure[Hurricane (after removing houses with positive $\rho$).]{
				{\includegraphics[width=0.495\textwidth,height=0.30\textwidth]{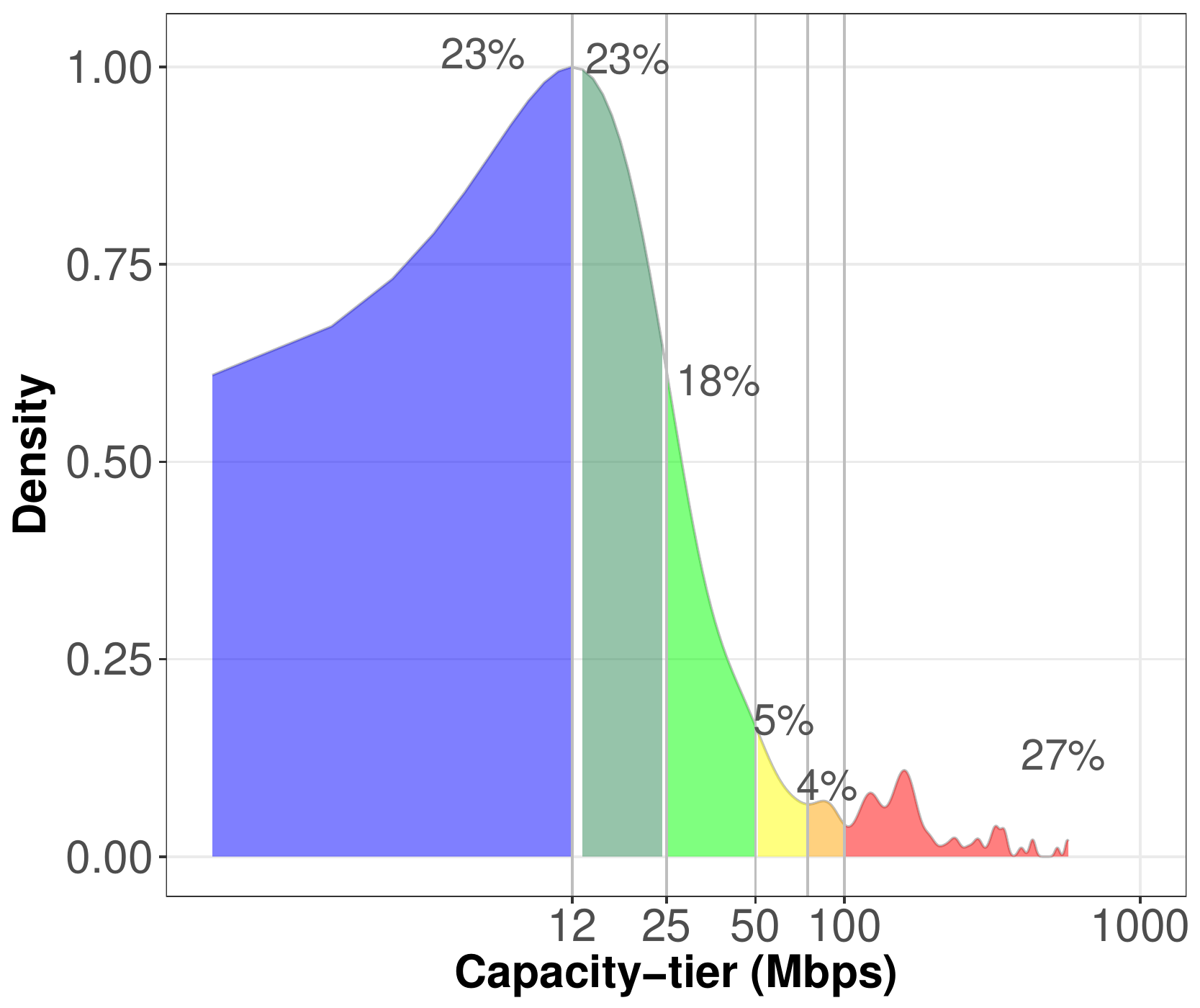}}\quad
				\label{fig:Hurricanebefore}
			}
		}
		
		\mbox{
			\subfigure[AT\&T (after eliminating outliers).]{
				{\includegraphics[width=0.495\textwidth,height=0.30\textwidth]{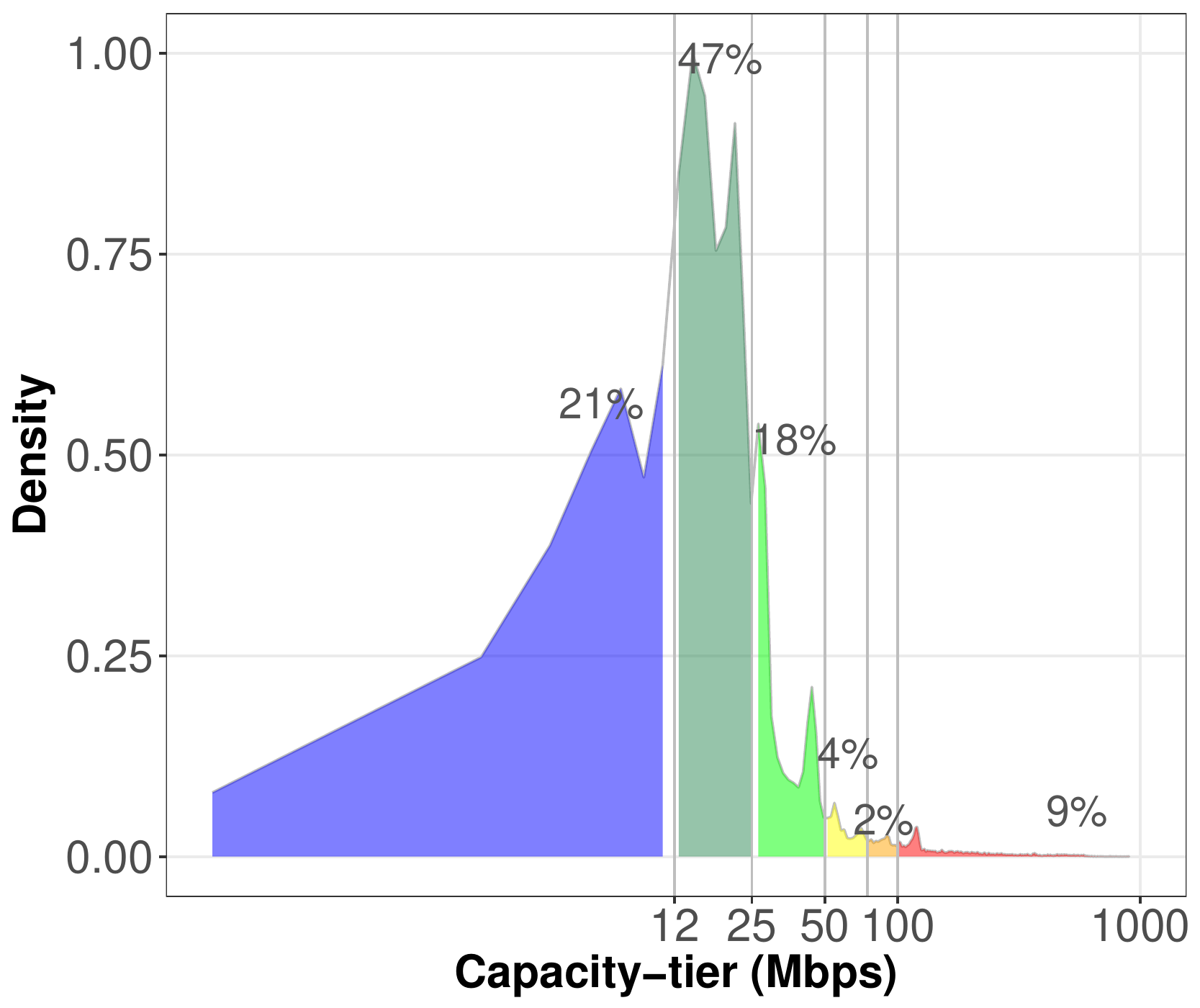}}\quad
				\label{fig:ATTafter}
			}
			\hspace{-2mm}
			\subfigure[Hurricane (after eliminating outliers).]{
				{\includegraphics[width=0.495\textwidth,height=0.30\textwidth]{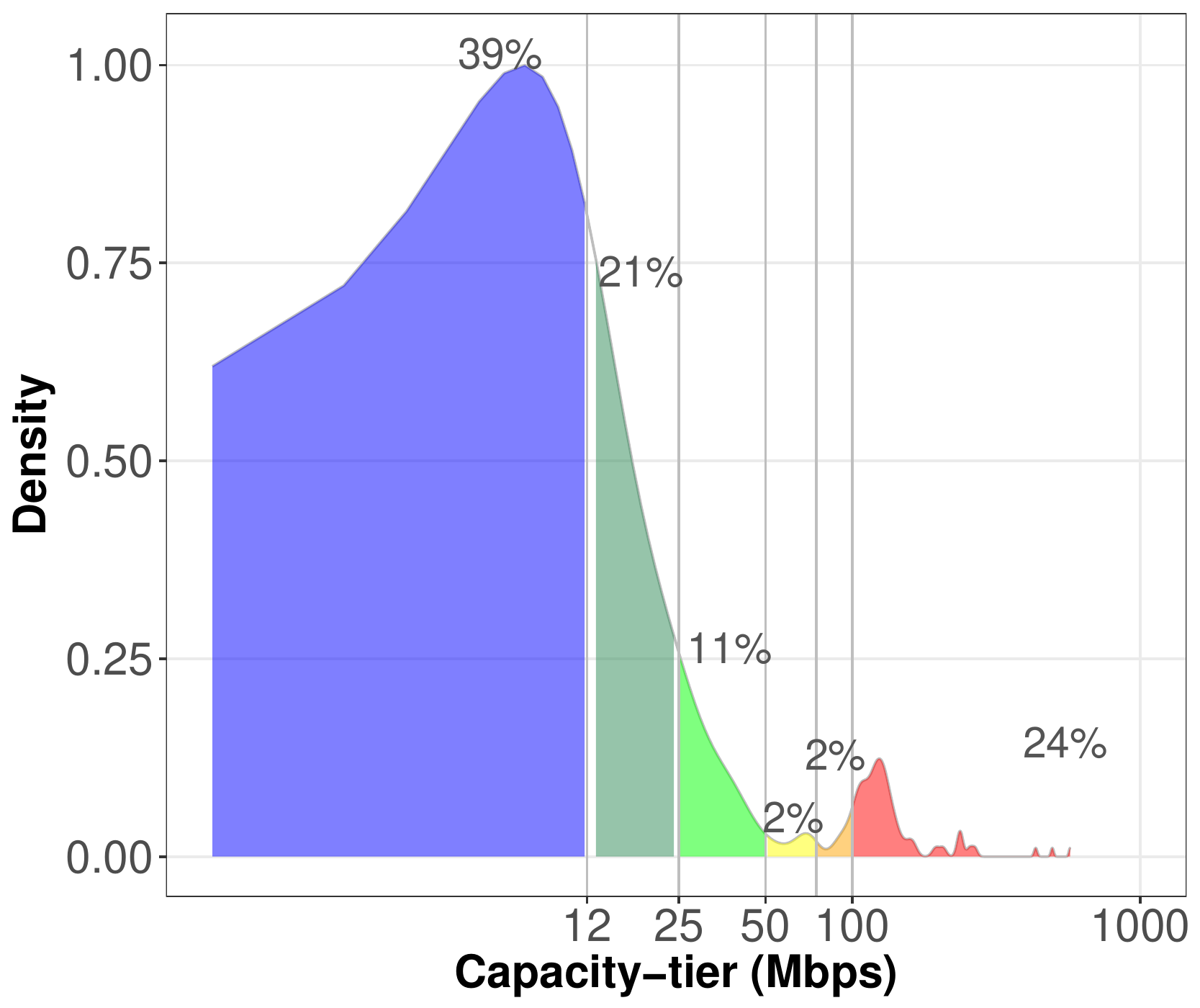}}\quad
				\label{fig:Hurricaneafter}
			}
		}
		\vspace{-5mm}
		\caption{Distribution of capacity tiers for (a,c,e) AT\&T and (b,d,e) Hurricane in US.}
		\vspace{-5mm}
		\label{fig:capISPUS}
	\end{center}
\end{figure*}

\begin{figure*}[t!]
	\begin{center}
		\mbox{
			\subfigure[Telstra (raw data -- each IP treated as a house).]{
				{\includegraphics[width=0.495\textwidth,height=0.30\textwidth]{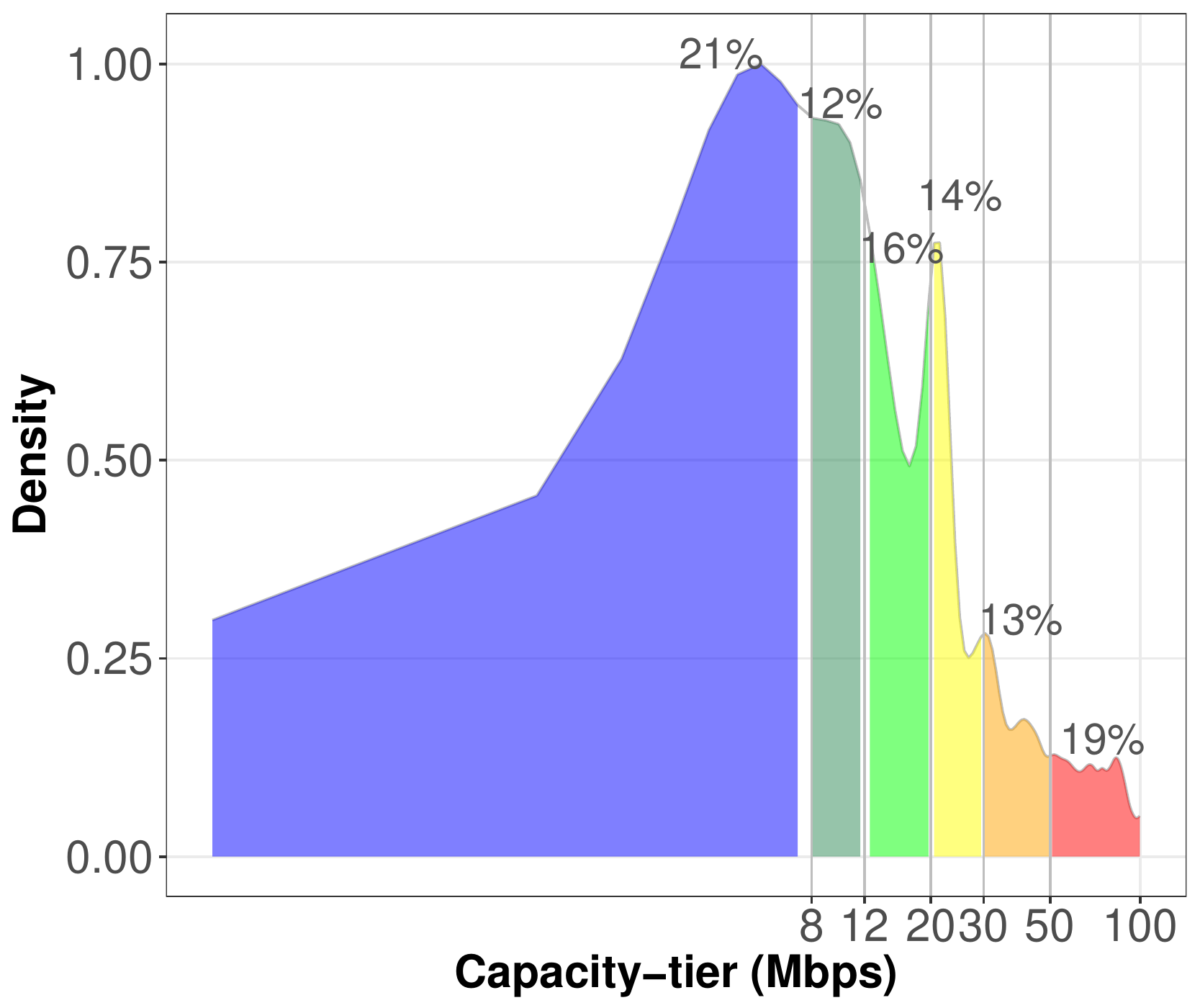}}\quad
				\label{fig:Telstraraw}
			}
			\hspace{-2mm}
			\subfigure[Optus (raw data -- each IP treated as a house).]{
				{\includegraphics[width=0.495\textwidth,height=0.30\textwidth]{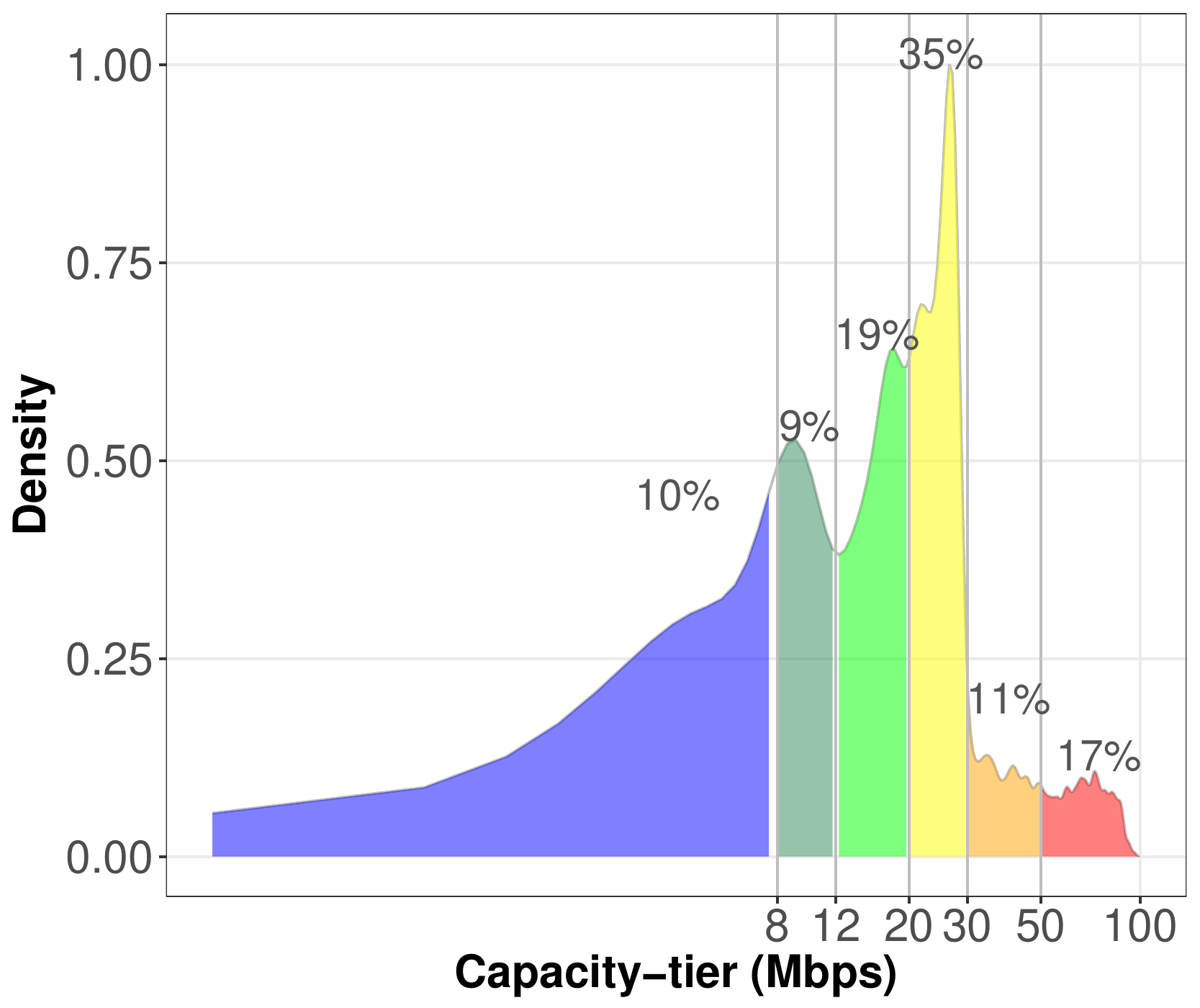}}\quad
				\label{fig:Optusraw}
			}
		}

		\mbox{
			\subfigure[Telstra (after removing houses with positive $\rho$).]{
				{\includegraphics[width=0.495\textwidth,height=0.30\textwidth]{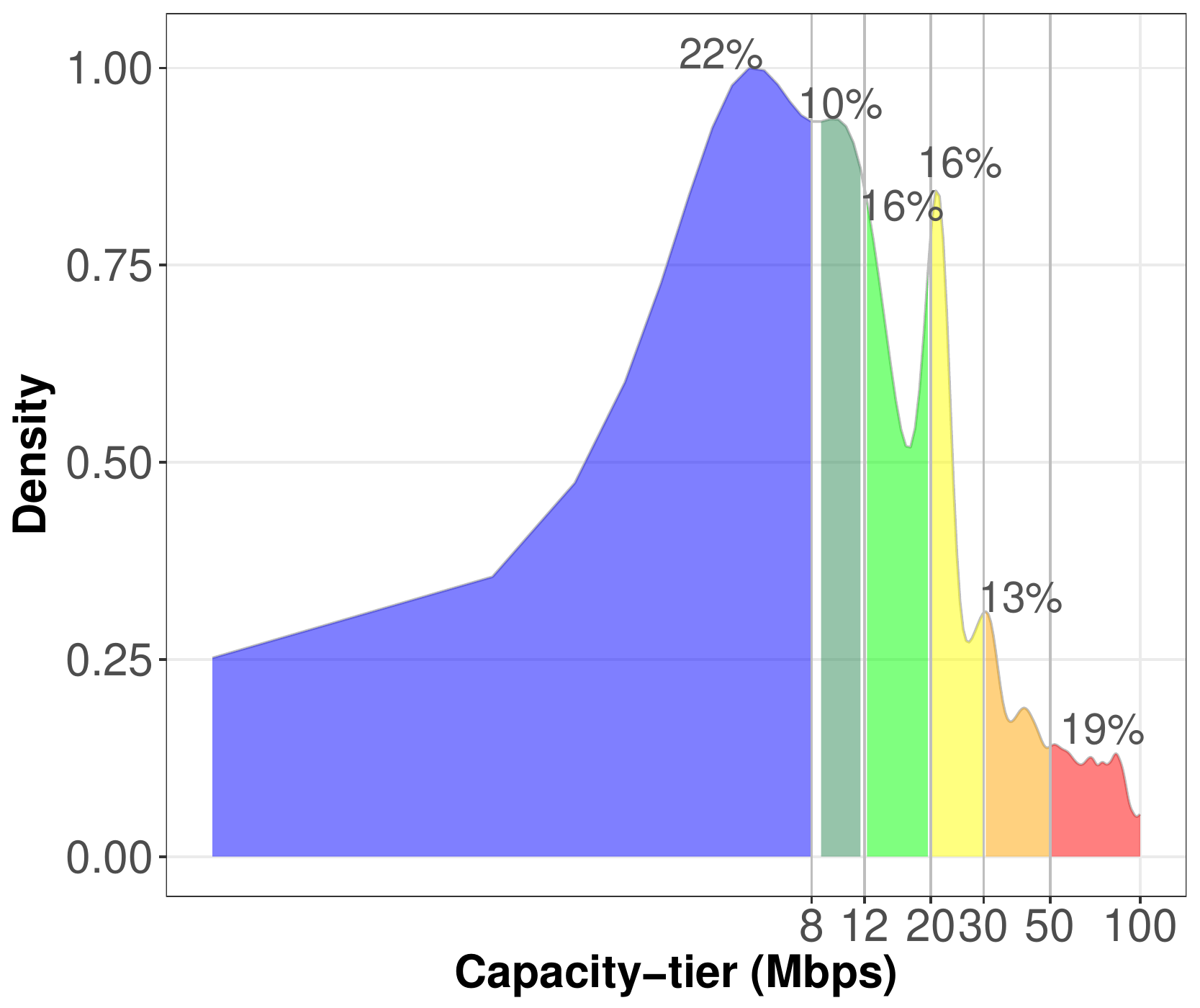}}\quad
				\label{fig:Telstrabefore}
			}
			\hspace{-2mm}
			\subfigure[Optus (after removing houses with positive $\rho$).]{
				{\includegraphics[width=0.495\textwidth,height=0.30\textwidth]{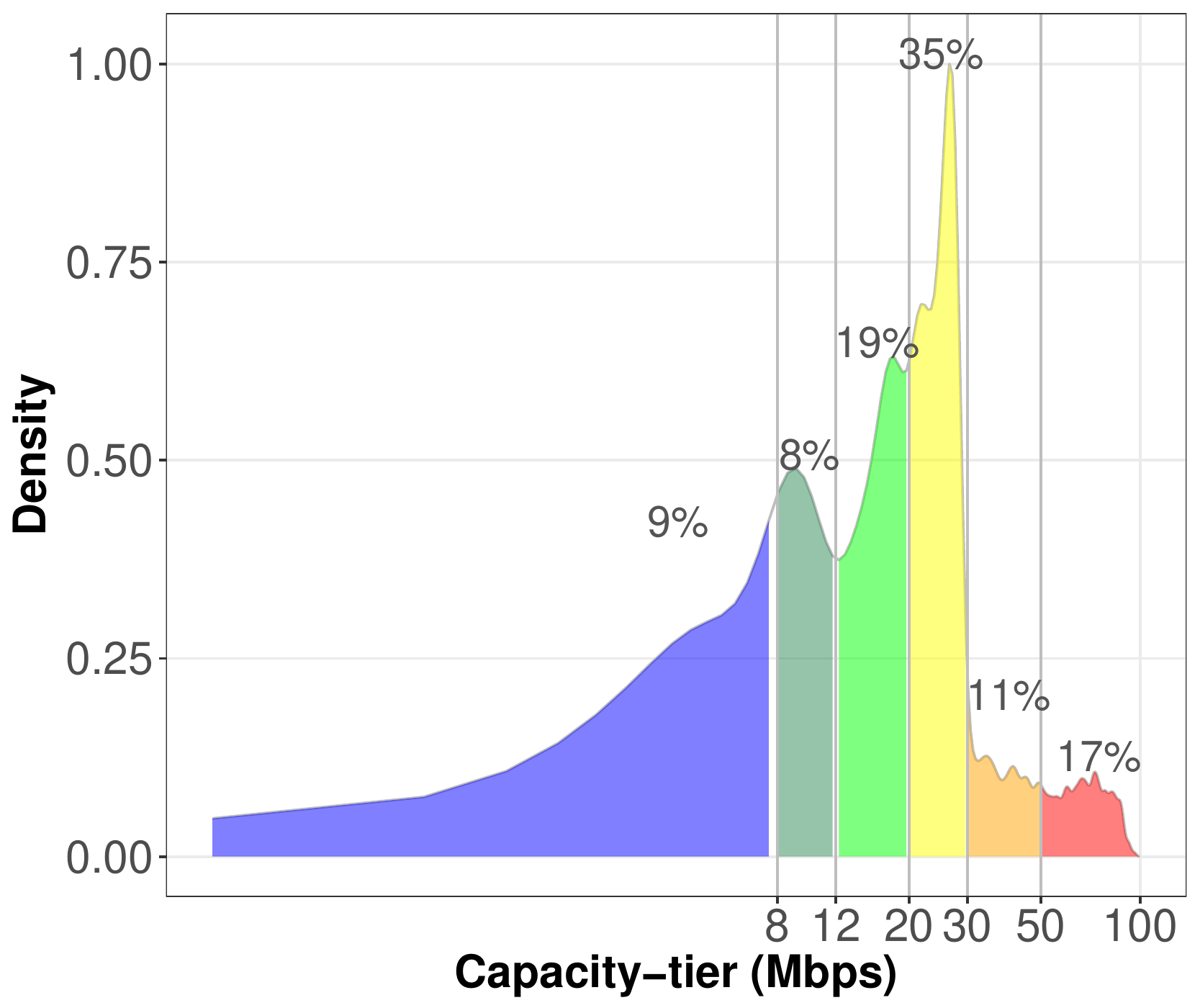}}\quad
				\label{fig:Optusbefore}
			}
		}
		
		\mbox{
			\subfigure[Telstra (after eliminating outliers).]{
				{\includegraphics[width=0.495\textwidth,height=0.30\textwidth]{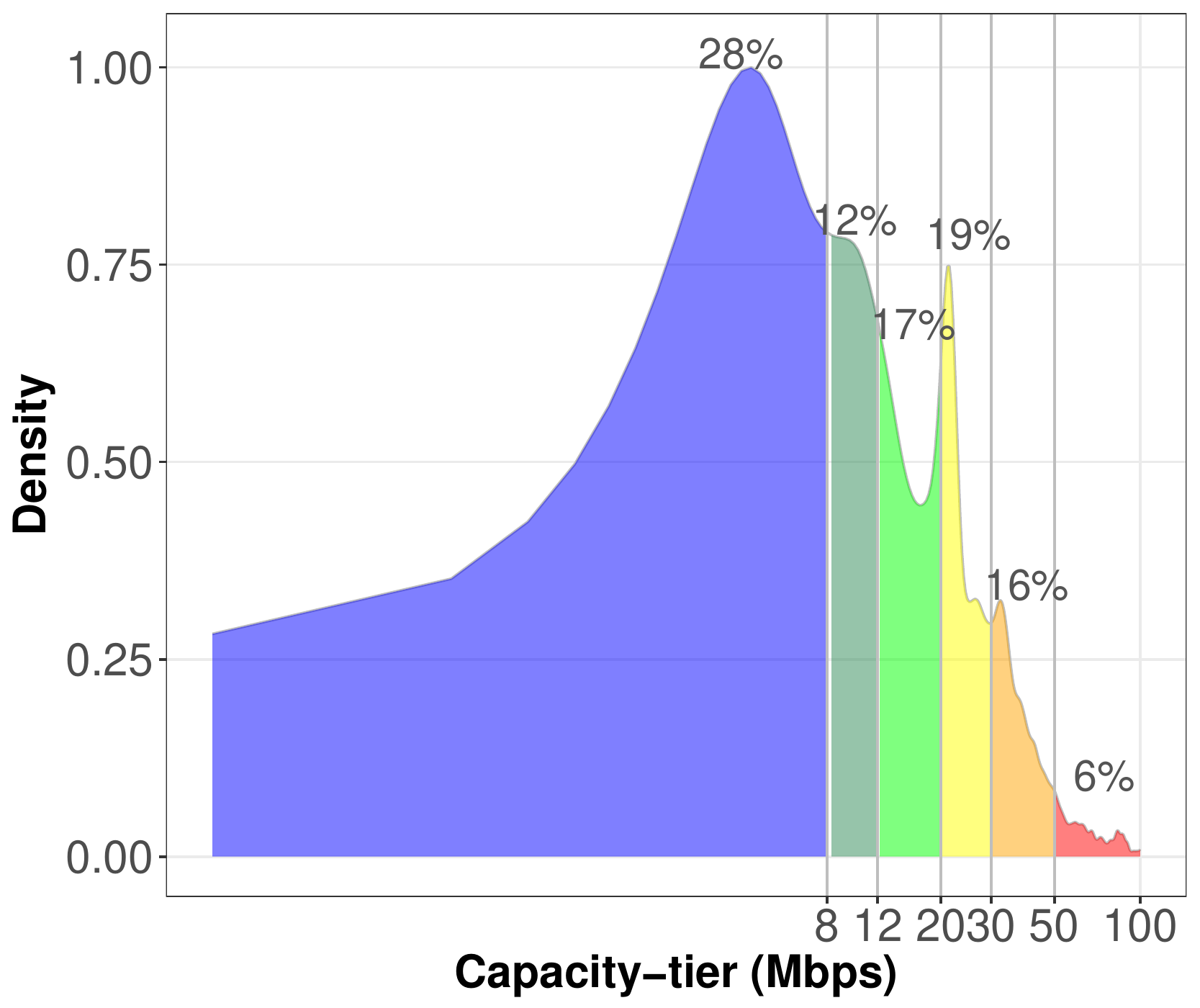}}\quad
				\label{fig:Telstraafter}
			}
			\hspace{-2mm}
			\subfigure[Optus (after eliminating outliers).]{
				{\includegraphics[width=0.495\textwidth,height=0.30\textwidth]{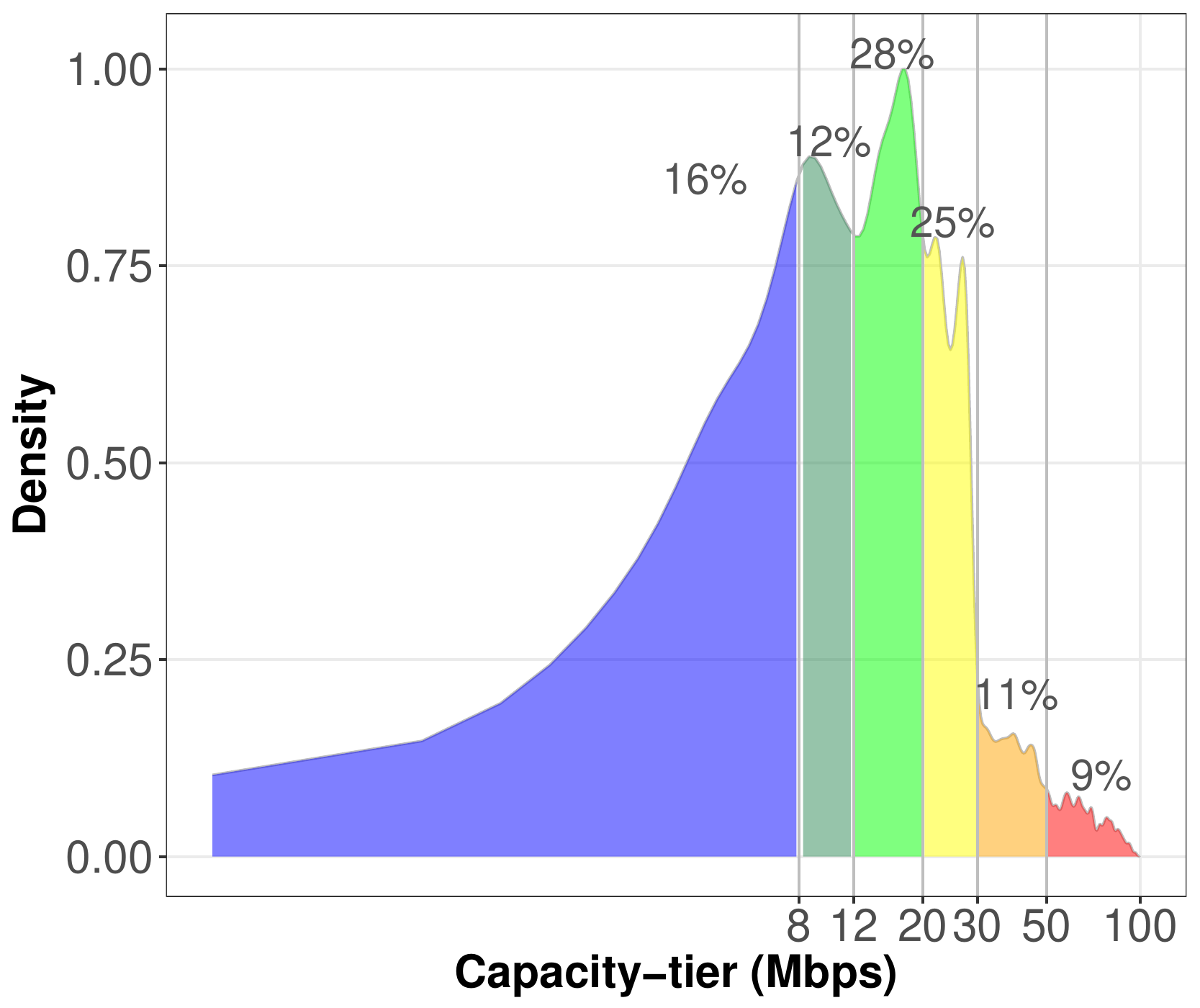}}\quad
				\label{fig:Optusafter}
			}
		}
		\vspace{-5mm}
		\caption{Distribution of capacity tiers for (a,c,e) Telstra and (b,d,e) Optus in AU.}
		\vspace{-5mm}
		\label{fig:capISPAU}
	\end{center}
\end{figure*}

\section{Estimating Household Speed-Tier}\label{sec:insights}
In previous section, we validated that the $\rho$ value is the key parameter associating measured data-points to households.
We therefore filter measurements corresponding to those IP addresses that exhibit positive correlation between their download-speed and congestion-count (i.e. $\rho > 0$). Fig.~\ref{fig:FilterSmapleCount} shows the number of IP addresses for each ISP (in AU and US) that are marked as single or multiple households using the $\rho$ value.  It is seen that for both countries,  IP addresses in large networks are predominantly mapped to single households shown by green bars -- $92.38$\% and $97.57$\% for Telstra and Optus respectively in AU, and $92.08$\% and $93.34$\% for AT\&T and Comcast respectively in US. On the other hand, a large fraction of IP addresses from small ISPs are filtered (shown by red bars) due to positive $\rho$ value. For example, no data from CEnet (in AU) is considered as single house.  

After removing data of multiple households, we now want to estimate the \textit{speed-tier} as a proxy for broadband capacity of each house. The download speed for a household will be limited by the capacity of its access link, which in turn is dictated by physical attributes such as medium (fiber, copper, wireless) and distance from the local exchange. It may further be constrained if the subscriber has chosen a plan with lower advertised speed. We term this maximum possible speed available to the household as its ``speed-tier''. As we will see later in this section, this attribute is
important when comparing ISPs, but is not explicitly present in the data since M-Lab is not privy to advertised speeds and subscriber plans. We therefore have to infer a household's access speed-tier from the measured data. We take the sensible approach of using the largest value of measured speed as the speed-tier for that household. 

As far as maximum download speed is concerned, in some cases we observe very large values in measurements which are more likely to be outliers. Fig.~\ref{fig:outliers} exemplifies measured download speed from three IP addresses. We use green solid lines to show the density distribution of speed overlayed by black circles stacked along the x-axis representing actual data points. We can see that there are several outliers observed around 60Mbps in   Fig.~\ref{fig:outliersMany} whereas the rest of measurements fall under 30Mbps -- the maximum speed value seems to be about half of outliers value. The dashed vertical red line depicts the cut-off point to filter outliers. Similarly, one data point in Fig.~\ref{fig:outliersOne} stretches the measured download speed to slightly over 50Mbps -- filtering this noise measurement results the speed-tier for address \textit{73.65.112.38} to be estimated as 20Mbps. We note that most of addresses contain a consistent clusters of measured speed, as shown in Fig.~\ref{fig:outliersZero}, thus no data point is discarded. 

\subsection{Outliers Detection}
In order detect and exclude outliers data in our study, we employ the standard modified Thompson Tau technique \cite{Tau} to statistically determine rejection zone. This method eliminates outliers more than two standard deviations from the mean value. After filtering outliers from our dataset, we pick the maximum value of remaining data points as the estimated speed-tier of corresponding house (i.e. IP address). Note that during outliers detection/filtering process, we compute a parameter called \textit{stretch factor} that we define as the ratio of maximum value of measured speed (including outliers) and the speed-tier (i.e. the maximum value of measured speed excluding outliers) for each house. Note that the stretch factor is always equal to or greater than 1. We show the CCDF plot of stretch factor in Fig.~\ref{fig:ccdf}. Interestingly, no outliers (i.e. the stretch factor equals 1) are detected in $36$\% and $35$\% of AU and US houses respectively. It is also important to note that majority of houses (i.e. $88$\% in AU and $85$\% in US) have the stretch factor less than 2. Unsurprisingly, large stretch factor (i.e. more than 5) are observed rarely, only $1\%$ of houses for both AU and US. We will see later that the distribution of stretch factor will affect our interpretation of capacity tiers when comparing various ISPs.

\subsection{Insights}  
We have so far applied data filtering at two stages: (a) discarding data shared by multiple houses (i.e. corresponding to positive $\rho$ values), and (b) eliminating unusual large measured speeds (i.e. outliers) from data of some houses. We believe that these two filters are crucial in removing irrelevant and noisy data points that allows meaningful analysis of performance tests at large scale when comparing ISPs by the metric of residential broadband capacity/speed tiers.

Lastly, our aim is to gain insights into capacity tiers across ISPs. We first create capacity bins of most commonly advertised tiers for each country and then plot the normalized density distribution of speed-tier for three scenarios:  (a) original raw NDT data, (b) processed data after isolating single houses, and (c) cleaned data after eliminating outliers, in Fig.~\ref{fig:capISPUS} for US and in Fig.~\ref{fig:capISPAU} for AU. 
The top row plots in both Figures~\ref{fig:capISPUS} and~\ref{fig:capISPAU} show capacity tiers when the original raw NDT dataset is used, the middle row corresponds to results when single houses are isolated, and the last row depicts the final results when our method is fully applied (single houses isolated and outliers eliminated). For US, we compare the capacity tiers of a large ISP (AT\&T) versus a small one (Hurricane), and for AU, we infer the landscape of capacity-tiers in two large Internet providers, i.e. Telstra and Optus.  

Obviously, the capacity distribution of large ISPs in both AU and US are less sensitive to our first stage of filtering (i.e. discarding data of multiple houses) and thus contribution of various bins changes marginally for: AT\&T in Fig.~\ref{fig:capISPUS}, and Telstra/Optus in Fig.~\ref{fig:capISPAU}. On the other hand, for Hurricane in Fig.~\ref{fig:capISPUS}, we see a significant change in two bins namely low-to-middle tier [12, 25] Mbps (i.e. $7$\% drop), and high tier [100,100+] Mbps (i.e. $11$\% rise) which is expected. 

Applying the outliers filter, however, significantly impacts the distribution of capacity tiers for all ISPs, large and small. For example, for AT\&T in Figures~\ref{fig:ATTbefore} and~\ref{fig:ATTafter}, two bins of low tier [0, 8] Mbps and low-to-middle tier [12, 25]Mbps see a rise of $5$\% and $7$\% respectively whereas the middle tier [25, 50]Mbps sees a $8$\% drop -- meaning that middle tier houses have more outliers eliminated, thus shifted to left inflating the contribution of lower bins. Shifting to lower tiers is more visible in plots for Hurricane (Figures~\ref{fig:Hurricanebefore} and~\ref{fig:Hurricaneafter}).

It is not surprising to spot ISPs are having very different speed-tier distribution plots in countries that each ISP owns its own broadband infrastructure. Even with a shared/nationalized broadband infrastructure countries like Australia, some ISPs may be serving customers with more lower speed tiers, which can drag their averages down. In fact, in AU, Telstra claims that it serves more rural/regional customers than other ISPs such as Optus, which is used as an argument why it ranks lower on the Netflix ISP speed index \cite{Netflix} constantly. When we compare speed-tier distribution after use of our methods as shown in Fig.~\ref{fig:Telstraafter} and Fig.~\ref{fig:Optusafter}, we see interestingly that Telstra does serve a large fraction of subscribers with the lowest speed tier (i.e. [0, 8]Mbps), more than Optus does ($28$\% vs. $16$\%). 
%In addition, Telstra has less fraction of medium speed tier ([12Mbps-20Mbps]) compared to Optus ($17$\% vs. $28$\%). It is also worth mentioning that if we compare Oputs raw data and cleaned data in Fig.~\ref{fig:Optusraw} and Fig.~\ref{fig:Optusafter}, we see that raw data shows seemly unrealistic low percentage of lowest speed-tier ([0Mbps-8Mbps]) than cleaned data ($10$\% vs. $16$\%).

%\begin{figure*}[t!]
%	\begin{center}
%		\mbox{
%			\subfigure[Number of filtered ATT]{
%				{\includegraphics[width=0.3\textwidth,height=0.20\textheight]{Fig/capISP/ATT_SpeedTier_changes}}\quad
%				\label{fig:ATT-FILTER}
%			}
%			\subfigure[Number of filtered Comcast]{
%				{\includegraphics[width=0.3\textwidth,height=0.20\textheight]{Fig/capISP/Comcast_SpeedTier_changes}}\quad
%				\label{fig:COMCAST-FILTER}
%			}
%			\hspace{0mm}
%			\subfigure[Number of filtered Hurricane]{
%				{\includegraphics[width=0.3\textwidth,height=0.20\textheight]{Fig/capISP/Hurricane_SpeedTier_changes}}\quad
%				\label{fig:Hurricane-FILTER}
%			}
%			\hspace{0mm}
%			
%		}
%		\vspace{-5mm}
%		\caption{By applying isolating households methods, the labeled multiple houses around low capacity tiers. for both (a,b). }
%		\vspace{-5mm}
%		\label{fig:HURRICANE-FILTER}
%	\end{center}
%\end{figure*}

\section{Conclusion}\label{sec:concl}
This paper has proposed systematic method for analysis of large scale performance data from M-Lab across broadband ISPs.
Our method is the first to isolate measurements from single households, remove noise data, and extract their broadband capacity. 
We first identified the correlation between measured speed and congestion count in M-Lab NDT data as the key indicator whether or not the client IP address represents a single house or multiple houses.
We then validate our hypothesis at small scale by analyzing ground truth data collected from two known houses, and at large scale by consistency check across ISPs of various size and across months. Finally, we apply our filtering method on data from selected network operators in Australia and the US, estimate the residential broadband capacity in their networks, and reveal insights into distribution of capacity tiers across ISPs.

\bibliographystyle{IEEEtran}
\bibliography{estimateCap}

\end{document}